\documentclass[aps,prb,twocolumn,groupedaddress,showpacs,superscriptaddress,amssymb,amsmath]{revtex4-1}
\usepackage{graphicx}
\usepackage{amsmath}
\usepackage{amssymb}
\usepackage{amsfonts}
\usepackage{color}
\usepackage{bm}        
\usepackage{comment}
\usepackage{verbatim}
\usepackage{placeins}
\usepackage[mathscr]{eucal}
\usepackage[colorlinks, linkcolor=red,citecolor=blue,urlcolor=blue]{hyperref}
\usepackage[normalem]{ulem}
\usepackage[all]{hypcap}
\usepackage{multirow}
\usepackage[dvipsnames,usenames,table]{xcolor}
\usepackage{dcolumn}
\usepackage{hyperref}
\usepackage{bm}
\usepackage{epsf}
\usepackage{subfigure}
\usepackage{amsfonts}
\usepackage{epstopdf}%
\usepackage{cancel}

\newcommand{\be}{\begin{equation}}
\newcommand{\ee}{\end{equation}}
\newcommand{\bea}{\begin{eqnarray}}
\newcommand{\eea}{\end{eqnarray}}
\def\langled{{\langle \langle}}
\def\rangled{{\rangle \rangle}}

\begin{document}
\title{Bounds on fluctuations for ensembles of quantum thermal machines}  
 
 \author{Matthew Gerry}
\altaffiliation{These authors contributed equally to this work}
	\affiliation{Department of Physics, University of Toronto, Toronto, Ontario, Canada M5S 1A7}
\author{Na'im Kalantar}	
\altaffiliation{These authors contributed equally to this work}
\affiliation{Department of Chemistry and Centre for Quantum Information and Quantum Control,
University of Toronto, 80 Saint George St., Toronto, Ontario, M5S 3H6, Canada}	
	\author{Dvira Segal}
\affiliation{Department of Chemistry and Centre for Quantum Information and Quantum Control,
University of Toronto, 80 Saint George St., Toronto, Ontario, M5S 3H6, Canada}
\affiliation{Department of Physics, University of Toronto, Toronto, Ontario, Canada M5S 1A7}
		\email{dvira.segal@utoronto.ca}
\date{\today}

\begin{abstract}
We study universal aspects of fluctuations in an ensemble of noninteracting continuous quantum thermal machines in the steady state limit.
Considering an individual machine, such as a refrigerator, in which relative fluctuations (and high order cumulants) of the cooling heat current to the absorbed heat current, $\eta^{(n)}$, are upper-bounded, $\eta^{(n)}\leq \eta_C^n$ with  $n\geq 2$ and $\eta_C$ the Carnot efficiency,
we prove that an {\it ensemble} of $N$ distinct machines similarly satisfies this upper bound on the relative fluctuations of the ensemble, $\eta_N^{(n)}\leq \eta_C^n$.
For an ensemble of distinct quantum {\it refrigerators}  with components operating in the tight coupling limit we further prove 
the existence of a {\it lower bound} on $\eta_N^{(n)}$ in specific cases,  exemplified on three-level quantum absorption refrigerators and resonant-energy thermoelectric junctions.
Beyond special cases, the existence of a lower bound on $\eta_N^{(2)}$ for an ensemble of quantum refrigerators is demonstrated by numerical simulations.
\end{abstract}

\maketitle 

%
\section{Introduction}

Significant efforts in stochastic and quantum thermodynamics 
\cite{st-thermo1, st-thermo2, st-thermo-Broeck,Q-thermo1,Q-thermo2} are currently devoted to understanding 
tradeoff relations in nanoscale thermal machines by weighting currents, their fluctuations, entropy production, and efficiency.
As an example, the ``Thermodynamic Uncertainty Relation" (TUR) describes a tradeoff between precision
(relative fluctuation) and cost (entropy production). Refs.  
\cite{Barato:2015:UncRel, trade-off-engine, Gingrich:2016:TUR, Horowitz:2017:TUR, Falasco, Garrahan18, 
Dechant:2018:TUR,Timpanaro, Bijay-TUR,Saito-TUR,Hasegawa,Miller,Junjie-TUR, Agarwalla-TUR}
constitute representative examples in this broad and active field.
The TUR further constrains the performance of thermal engines, balancing
output power, power fluctuations and the engine's efficiency \cite{trade-off-engine}.

A separate class of bounds, which are independent of the TUR, concerns ratios between fluctuations 
of different currents: the output power and input heat current \cite{Watanabe,Sagawa,Gerry, Bijay1, Bijay2}.
For the process of refrigeration one considers the ratio 
\bea
\eta^{(n)} \equiv \frac{\langle  \langle q^n\rangle \rangle }{\langle \langle w^n\rangle \rangle},
\label{eq:etanI}
\eea
with $q$ as the extracted stochastic cooling heat current and $w$  the corresponding input heat current;
 $\langled A^n\rangled$, $n=1,2,...$ is the $n$th cumulant of a stochastic variable $A$.
For $n=2$ this ratio is lower and upper-bounded in linear response for continuous machines in steady state under time reversal symmetry \cite{Gerry},  
\bea
 \eta ^2\leq\eta^{(2)}\leq \eta_C^2.
\label{eq:eta2I}
\eea
$\eta_C$ is the Carnot bound and $ \eta $ (our short notation for $\eta^{(1)}$) stands for the efficiency of the machine.
Based on the upper bound on $\eta^{(2)}$, a tighter-than-Carnot efficiency bound had been derived \cite{Watanabe,Sagawa,Gerry,Bijay1}, tighter than bounds received from the
TUR \cite{Bijay2}. 

In this paper, we focus on an {\it ensemble} of {\it distinct}, continuous,  steady-state thermal machines, and inquire on universal aspects of their fluctuations. The machines operate between the same affinities (temperatures, chemical potentials), but are different in their working parameters such as internal energies, system-bath coupling strength. 
We then ask the following basic question: Assuming the relative fluctuations of an {\it individual} machine are upper and lower bounded according to Eq. (\ref{eq:eta2I}), do these bounds hold for an ensemble of distinct, uncorrelated machines? 

This question is not trivial, as we show below. It is particularly difficult to address the lower bound: 
While all our machines are upper-bounded by the same Carnot limit, their efficiencies $\eta$ (dictating the lower bound) are distinct. 
To make progress, we consider here individual machines operating in the so-called tight coupling (TC) limit \cite{Caplan,benenti17}: In each machine, the stochastic input and output currents are proportional to each other, $\omega\propto q$.
Under the TC limit, one can readily prove, beyond linear response, the validity of the upper bound in Eq. (\ref{eq:eta2I}), and show that the lower bound is saturated---for an individual machine \cite{Gerry}.
However, tight-coupling does not necessarily hold when studying a collection of subsystems,
even when the constituent elements follow it.
 Therefore, as we discuss here, proving in general the lower bound on $\eta^{(2)}$ is a nontrivial task.
 
 We now pose the problem to be addressed in this study. 
 We consider {\it individual} TC machines (e.g., refrigerators) whose
relative fluctuations $\eta_k^{(n)}$ [see Eq. (\ref{eq:etanI})] for each member $k$ are bounded according to
 \bea
 \eta_k^n = \eta_k^{(n)} \leq \eta_C^{n}, \,\,\,\,\,\ n=1,2,...
 \label{eq:etaTC}
 \eea
 with the efficiency $\eta_k$ possibly different for each member in the ensemble.
Our objective is to find whether an {\it ensemble} of $N>1$ distinct, independent machines satisfies the relations
 \bea
 \eta_N^n \leq \eta_N^{(n)} \leq \eta_C^{n}, \,\,\,\,\,\ n=2,3,...
 \label{eq:goal}
 \eea
arbitrarily far from equilibrium; the validity of the upper and lower bounds in the linear response regime was proved in Ref. \cite{Gerry}.
The currents considered in devising the ratios (\ref{eq:etanI}) are the total ones,
$w=\sum_{k=1}^Nw_k$ and $q=\sum_{k=1}^Nq_k$.
That is, e.g. $\eta_N^{(n)}$ is the efficiency ($n$=1) or relative fluctuations ($n=2$) of a system made of $N$ constituents. 
Expressing Eq. (\ref{eq:goal}) in words:
Considering ratios of cumulants of output to input currents in a collection of thermal machines,
we would like to show that this ratio is upper bounded---by the $n$th power of the Carnot efficiency.
Further, we interrogate whether a lower bound holds for the ensemble, given by the ensemble-averaged efficiency to the power $n$.

As we show in this paper, the upper bound in Eq. (\ref{eq:goal}) can be readily proved for any order $n$. 
As for the lower bound, we mostly restrict ourselves to the behavior of fluctuations, $n=2$, and we prove it in specific limits.
Beyond those cases, we establish the lower bound more broadly for refrigerators based on numerical simulations of three-level quantum absorption refrigerators (3LQARs) and thermoelectric junctions.
In contrast, we find examples of pairs ($N=2$) of thermoelectric {\it engines} that violate the lower bound in Eq. (\ref{eq:goal})  in the far from equilibrium regime.
  
%

The question posed here, on the validity of bounds for an ensemble of machines, is critical to our ability to experimentally confirm fundamental theoretical results.
In some experiments, quantum machines are constructed from a collection of systems, such as trapped ions \cite{Poem}, quantum dots (see Review \cite{QD-rev}), and molecules  \cite{engine-expt-spin}.
Practically, even when aiming for homogeneity, these components cannot be made precisely identical in their energies and couplings to the environment.
Moreover, even for machines made from an individual `particle' \cite{singleatom,LinkeE,ionE,engine-expt-spin-osc,Widera},
experiments inevitably suffer from a certain degree of uncertainty, requiring an ensemble average.
Considering e.g. quantum absorption refrigerators based on superconducting circuits \cite{Hofer} or  
nitrogen vacancy centers in diamond \cite{bargil}, internal energies defining the refrigerator, as well as
system-bath coupling parameters cannot be precisely-repeatedly realized, thus measurements necessarily rely on averaging. 

We highlight that the ratio of cumulants as defined in Eq. (\ref{eq:etanI}) is unrelated to the concept of efficiency fluctuations explored e.g. in Refs. \cite{effF1,effF2,effF3,Galperin,Denzler}. In these studies, one defines the stochastic efficiency and studies its probability distribution function. In contrast, here we focus on the currents as the 
stochastic variables, we construct their cumulants, then look at their ratios to define $\eta^{(n)}$.

As a final introductory comment, 
while our examples concern quantum thermal machines,
their quantum nature only involves the discreteness of their energy levels and the quantum statistics of the baths (bosonic or fermionic). As such, bounds discussed in this work are directly applicable to classical systems. 

The paper is organized as follows. 
The upper bound is proved in Sec. \ref{SecU}.
In Sec. \ref{SecL}, we discuss the lower bound and arrange it in alternative forms.
We investigate the validity of the lower bound in different limits with two models for refrigeretors: In Sec. \ref{SecQAR} we focus on 3LQARs, while in Sec. \ref{SecTE} we examine thermoelectric junctions. Appendices A, B, and C provide additional proofs and supporting simulations for the lower bound in different limits. In Appendix D we demonstrate that the lower bound can be violated for thermoelectric engines operating far from equilibrium.


\section{Universal upper bound on ratio of fluctuations}
\label{SecU}

In this Section, we prove that arbitrarily far from equilibrium,
ratios of cumulants of order $n$ of an ensemble with $N$ noninteracting and uncorrelated {\it distinct} heat machines, operating under the same affinities, are bounded by
\bea
\eta^{(n)}_{N}\leq \eta_C^n,
\label{eq:upper}
\eea
so long as the inequality is satisfied at the level of the individual machine, $\eta^{(n)}_{k}\leq \eta_C^n$.
$\eta_C$ is the Carnot bound dictated by the temperatures common to all machines.

The proof holds for engines and refrigerators; we describe it in the language of refrigerators.
We consider small, possibly quantum, refrigerators with $w=\sum_{k=1}^N w_k$ the total stochastic heat current absorbed in the cooling process from the so-called work bath
and $q=\sum_{k=1}^N q_k$ the total extracted heat current from the cold bath.
The components operate independently and are not correlated.
We assume that each individual member of the ensemble satisfies 
\bea
\eta_k&\equiv& \frac{\langle q_k \rangle }{\langle w_k \rangle} \leq \eta_C,
\nonumber\\
\eta_k^{(2)}&\equiv& \frac{\langled q_k^2 \rangled }{\langled w_k^2 \rangled} \leq \eta_C^2,
\nonumber\\
\eta_k^{(n)}&\equiv& \frac{\langled q_k^n \rangled }{\langled w_k^n \rangled} \leq \eta_C^n, \,\,\, n>2.
\label{eq:upperlist}
\eea
To assist readers, we explicitly included the first two definitions.
It can be shown that Eq. (\ref{eq:upperlist}) holds in the TC limit even far from equilibrium---for individual machines \cite{Gerry} .
 Concrete systems operating in the TC limit are 3LQARs and resonant-level thermoelectric junctions \cite{Gerry}.
 
For clarity, we begin with the case $N=2$ and consider two refrigerators $A$ and $B$.
It is not difficult to prove that the ratio of their total fluctuations are bounded by the Carnot-efficiency squared,
\begin{widetext}
\bea
\eta^{(n)}_{N=2}&\equiv&
\frac{\langled (q_A+ q_{B})^n\rangled}{\langled (w_A+w_B)^n\rangled} = 
\frac{\langled q_A^n\rangled +\langled q_{B}^n\rangled}{\langled w_A^n\rangled +\langled w_B^n\rangled}
\nonumber\\
&&
=\frac{\langled q_A^n\rangled }{\langled w_A^n\rangled}\left(1-   \frac{\langled w_B^n\rangled }{\langled w_A^n\rangled +\langled w_B^n\rangled } \right)
+\frac{\langled q_B^n\rangled }{\langled w_B^n\rangled}\left(1-   \frac{\langled w_A^n\rangled }{\langled w_A^n\rangled +\langled w_B ^n\rangled } \right)
\nonumber\\
&&\leq \eta_C^n\left(1-   \frac{\langled w_B^n\rangled }{\langled w_A^n\rangled +\langled w_B^n\rangled } \right)
+\eta_C^n\left(1-   \frac{\langled w_A^n\rangled }{\langled w_A^n\rangled +\langled w_B^n\rangled } \right)
=\eta_C^n.
\eea
\end{widetext}
%
In the first line, we use the fact that the machines are uncorrelated. The last line is arrived based on bounds for the individual machines. 

Next, along the same principle we prove by induction the $(N+1)$th inequality based on the validity of an upper bound for an $ N$-member ensemble. We denote by $q_k$ and $w_k$ the stochastic current of the k$th$ machine.
$q_{N+1}$ and $w_{N+1}$ are the stochastic currents of the $(N+1)$th member of the ensemble; $\eta_N$ (and similarly for higher $n$) is the efficiency of an $N$-sized ensemble. 
We now write
\bea
&&\eta^{(n)}_{N+1}\equiv \frac{\sum_{k=1}^{N+1} \langled q_k^n\rangled}{\sum_{k=1}^{N+1}\langled w_k^n\rangled }
\nonumber\\
&&=\frac{\sum_{k=1}^{N} \langled q_k^n\rangled +\langled q_{N+1}^n\rangled }{\sum_{k=1}^{N}\langled w_k^n\rangled  +\langled w_{N+1}^n\rangled }
\nonumber\\
&&=
\frac{\sum_{k=1}^{N} \langled q_k^n\rangled }{\sum_{k=1}^{N}\langled w_k^n\rangled   }
\left( 
1- \frac{\langled w_{N+1}^n\rangled}{\sum_{k=1}^{N} \langled w_k^n\rangled +\langled w_{N+1}^n\rangled }
\right)
\nonumber\\
&&+
\frac{ \langled q_{N+1}^n\rangled }{\langled w_{N+1}^n\rangled   }
\left( 
1- \frac{\sum_{k=1}^N\langled w_{k}^n\rangled}{\sum_{k=1}^{N} \langled w_k^n\rangled +\langled w_{N+1}^n\rangled }
\right)
\nonumber\\
&& \leq \eta_C^n.
\label{eq:etaNn}
\eea
In the last step we used $\frac{\sum_{k=1}^{N} \langled q_k^n\rangled }{\sum_{k=1}^{N}\langled w_k^n\rangled   }\leq \eta_C^n$
per our assumption of the validity of the upper bound for an ensemble with $N$ elements. We also made use of
$\frac{\langled q_{N+1}^n\rangled }{\langled w_{N+1}^n\rangled} \leq \eta_C^n$, valid for every individual machine. 

\begin{figure}[h]
{\includegraphics[width=7.7cm] {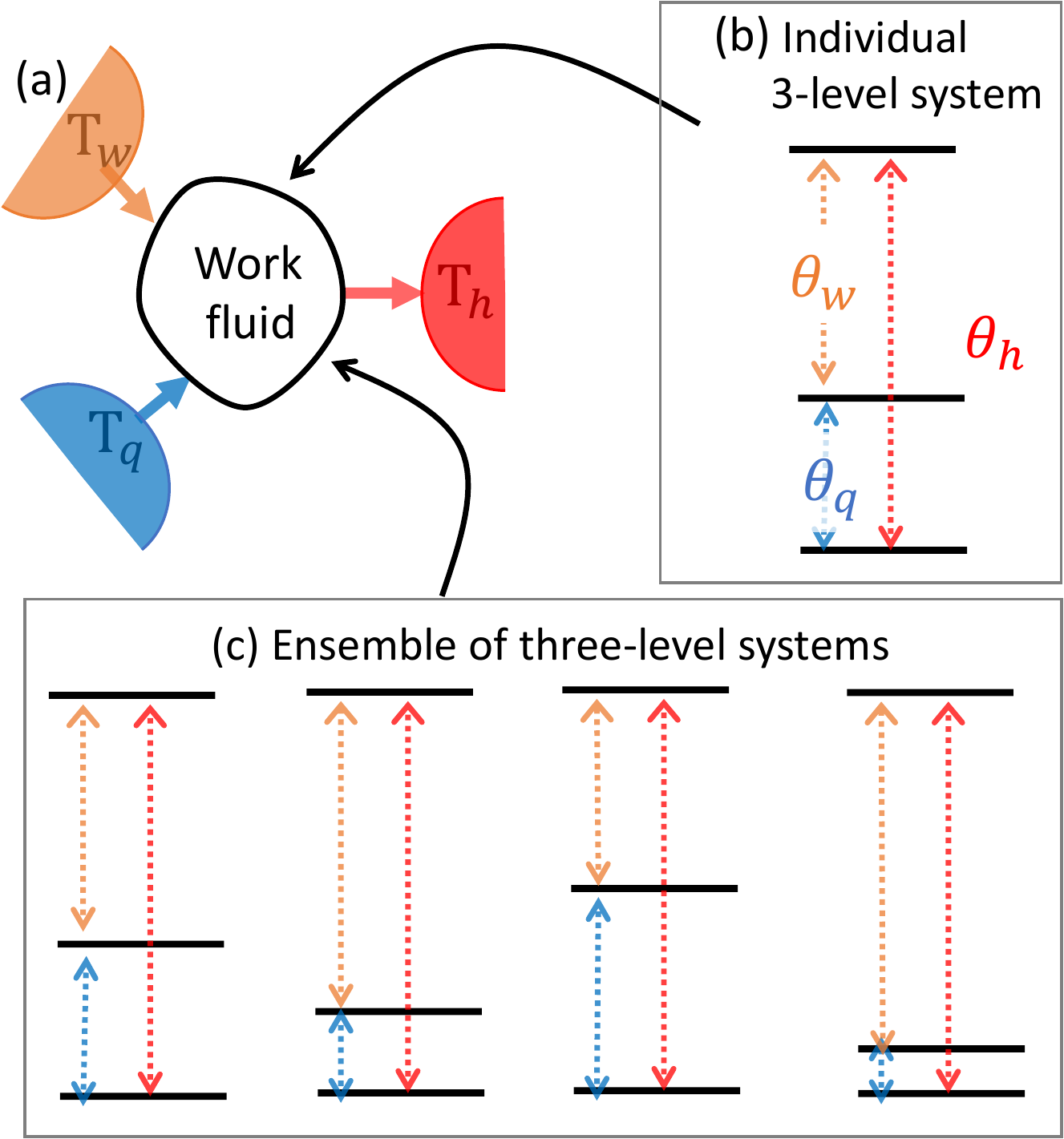} 
\caption{(a) Quantum absorption refrigerators
operating with (b) an individual 3-level system as its working fluid, (c) an ensemble of distinct 3-level systems, e.g.
of different energy spacing and distinct system-bath coupling parameters.
The temperatures of the heat baths are marked by $T_w>T_h>T_q$. 
The transition between the ground level to the intermediate one, of energy gap $\theta_q$ is coupled to the cold bath. 
The transition between the intermediate level to the top one, of gap $\theta_w$, is enacted by the work bath.
In a cooling operation, heat absorbed by the three-level system from the cold and work baths is emitted to the  hot  bath.
}}
\label{fig:QAR}
\end{figure}

Summing up, we proved that if the inequality $\eta_{N}^{(n)}\leq \eta_C^n$ holds for an individual machine, it also holds for a machine made of 
a collection of $N>1$  {\it distinct} systems---as long as they operate between the same temperatures, thus bounded by the same Carnot bound.
The working elements of our machine are made distinct in their internal parameters and their coupling to the surroundings.
For example,  in Fig.  \ref{fig:QAR} we depict a refreigerator with its working fluid including multiple three-level systems that are distinct in their
energy spacings. In Fig. \ref{fig:te_illustration}, we illustrate a thermoelectic device that comprises an array of independent junctions.

\section{Lower bound on ratio of fluctuations}\label{sec:lower_general}
\label{SecL}

Unlike the upper bound that we proved in Sec \ref{SecU}, we are not able to prove the lower bound in general, but only in specific cases (Sec. \ref{SecQAR} and Sec. \ref{SecTE}).
Before discussing these cases, in this section, we limit ourselves to a pair of systems ($N=2$) and organize the lower bound in a compatible, curious form.

We consider two machines, labelled $A$ and $B$, both operating in the tight-coupling limit thus satisfying the relation $\eta_k^2  = \eta_k^{(2)}$. 
We focus on the relation
\bea
\eta_{N=2}^2\leq\eta_{N=2}^{(2)}
\label{eq:low}
\eea
for the combined $N=2$ system. 
The ratio of fluctuations for the pair is given by
\bea
    \eta^{(2)}_{N=2}&=&
    \frac{\langled (q_A+ q_{B})^2\rangled}{\langled (w_A+w_B)^2\rangled} = 
    \frac{\langled q_A^2\rangled +\langled q_{B}^2\rangled}{\langled w_A^2\rangled +\langled w_B^2\rangled}
    \nonumber\\
    &=&\eta_A^{(2)}\frac{\langled w_A^2\rangled}{\langled w^2\rangled} + \eta_B^{(2)}\frac{\langled w_B^2\rangled}{\langled w^2\rangled}
    \nonumber\\
    &=&\lambda\eta_A^{(2)} + (1-\lambda)\eta_B^{(2)},
\eea
where $\lambda\equiv\langled w_A^2\rangled/\langled w^2\rangled$. Crucially, $0\leq\lambda\leq1$. If we assume without loss of generality that $\eta_A^{(2)}\geq\eta_B^{(2)}$, we have that $\eta_B^{(2)}\leq\eta_{N=2}^{(2)}\leq\eta_A^{(2)}$.

We may similarly expand the square of the efficiency itself,
\bea
    \eta_{N=2}^2&=&\frac{(\langle q_A\rangle + \langle q_B\rangle)^2}{\langle w\rangle^2}
    \nonumber\\
    &=&\eta_A^2\frac{\langle w_A\rangle^2}{\langle w\rangle^2} + \eta_B^2\frac{\langle w_B\rangle^2}{\langle w\rangle^2} + 2\eta_A\eta_B\frac{\langle w_A\rangle\langle w_B\rangle}{\langle w\rangle^2}
    \nonumber\\
    &=&\eta_A^{(2)}\frac{\langle w_A\rangle^2}{\langle w\rangle^2} + \eta_B^{(2)}\frac{\langle w_B\rangle^2}{\langle w\rangle^2} + 2\eta_A\eta_B\frac{\langle w_A\rangle\langle w_B\rangle}{\langle w\rangle^2},
    \nonumber
\eea
where in the last line we have used the strict equality for individual machines due to the tight-coupling limit. 
Because of the TC limit, the relation $\eta_A^{(2)}\geq\eta_B^{(2)}$ further implies that $\eta_A\geq \eta_B$, and $\eta_A\geq \sqrt{\eta_B^{(2)}}$. With these relations, we get
\bea
\eta^2_{N=2} &\leq& \eta_A^{(2)} \left[   
\frac{\langle w_A\rangle^2}{\langle w\rangle^2} + \frac{\langle w_B\rangle^2}{\langle w\rangle^2} + 2\frac{\langle w_A\rangle\langle w_B\rangle}{\langle w\rangle^2} \right] = \eta_A^{(2)},
\nonumber\\
\eta^2_{N=2} &\geq& \eta_B^{(2)} \left[   
\frac{\langle w_A\rangle^2}{\langle w\rangle^2} + \frac{\langle w_B\rangle^2}{\langle w\rangle^2} + 2\frac{\langle w_A\rangle\langle w_B\rangle}{\langle w\rangle^2} \right] = \eta_B^{(2)},
\nonumber
\eea
thus
$\eta_B^{(2)}\leq\eta_{N=2}^2\leq\eta_A^{(2)}$. 
It is now immediately clear that if $\eta^{(2)}_A = \eta_B^{(2)}$, a strict equality, $\eta_{N=2}^2=\eta_{N=2}^{(2)}$, is achieved. 
More generally, there exists some $0\leq\kappa\leq1$ such that $\eta_{N=2}^2 = \kappa\eta_A^{(2)} + (1-\kappa)\eta_B^{(2)}$. Solving for this coefficient gives
\be
    \kappa = \frac{\langle w_A\rangle^2}{\langle w\rangle^2} + 2\frac{\eta_B}{\eta_A + \eta_B}\frac{\langle w_A\rangle\langle w_B\rangle}{\langle w\rangle^2}.
\ee
The relation on the pair of machines, $\eta_{N=2}^2\leq\eta_{N=2}^{(2)}$, then, is satisfied exactly when $\kappa\leq\lambda$, or,
\be
    \frac{\langle w_A\rangle^2}{\langle w\rangle^2} + 2\frac{\eta_B}{\eta_A + \eta_B}\frac{\langle w_A\rangle\langle w_B\rangle}{\langle w\rangle^2} \leq \frac{\langled w_A^2\rangled}{\langled w^2\rangled}.
\label{eq:low2}
\ee
Since $2\eta_B/(\eta_A + \eta_B)\leq1$, this is always the case for pair of machines meeting the stronger condition, $\langled w_A^2\rangled/\langled w^2\rangled\geq\langle w_A\rangle/\langle w\rangle$, or, equivalently,
\be
    \frac{\langled w_A^2\rangled}{\langled w_B^2\rangled}\geq\frac{\langle w_A\rangle}{\langle w_B\rangle} ,\,\,\, {\rm for }\,\, \eta_A\geq \eta_B.
    \label{eq:low3}
\ee
Equation (\ref{eq:low2}) is equivalent to the lower bound (\ref{eq:low}).
In contrast, Eq. (\ref{eq:low3}) provides a stronger condition: Satisfying Eq. (\ref{eq:low3}) necessarily means obeying the original relation, (\ref{eq:low}).
However, we may violate the inequality (\ref{eq:low3}) yet still satisfy the lower bound (\ref{eq:low}).

.

\section{Model I: Ensemble of absorption refrigerators}
\label{SecQAR}

Fundamental results on autonomous-continuous thermal machines are often illustrated and examined within simple models for
quantum absorption refrigerators  \cite{Kosloff, Luis, MarkRev14}; 
recent experiments realized a QAR using trapped ions \cite{Poem,ionE}. 
In a QAR, heat is extracted from a cold ($q$) bath and released into a hot ($h$) bath by absorbing heat from the so-called work ($w$) reservoir.
The reversed operation realizes a heat engine. 
We identify three temperatures, $T_w>T_h>T_q$ in a QAR, and three stochastic heat currents, $w,h,q$, defined positive when flowing towards the  system.

The performance of QARs has been investigated in different models for elucidating principles in quantum thermodynamics. For example, QARs have been analyzed from weak to strong couplings to the baths \cite{Brandes,AMu, HavaNJP,Tanimura16}, in models supporting multiple competing cycles \cite{Correa15,HavaM}, and when quantum coherences between eigenstates survive in the steady state limit \cite{KilgourQAR,Junjie21}; these are illustrative examples out of a rich literature.
In this paper, we utilize the three-level model 
of Scovil and Schulz-DuBois \cite{Scovil} to illustrate bounds on relative fluctuations of currents for an ensemble of QARs. We limit our discussion to the weak system-bath coupling limit, which can be handled with a perturbative quantum Master equation,
providing the full counting statistics of the model \cite{Segal18,Junjie21}.

A schematic diagram of our model is displayed in Fig. \ref{fig:QAR}(a). 
An individual 3LQAR, illustrated in Fig. \ref{fig:QAR}(b), had particularly served to elucidate concepts in quantum thermodynamics, since at weak system-bath coupling it can be analytically solved. 
In this model, the three baths are coupled selectively to the different transitions: The cold ($q$) bath allows the transitions across $\theta_q$, from the ground state to the intermediate level. The work ($w$) bath is coupled to a transition of energy gap $\theta_w$, from the intermediate level to the top one. In a cooling process, the hot bath ($h$) extracts the heat, $\theta_q+\theta_w$.

\subsection{Single refrigerator, $N=1$}
Consider an individual 3LQAR of spacings $\theta_q$ and $\theta_w$.
The cooling efficiency of the engine is defined as
$\eta= \frac{\langle q\rangle}{\langle w \rangle}$ with $q$ ($w$) the stochastic cooling  (work) heat currents.
It can be shown that in the weak system-bath coupling limit, an individual  3LQAR operates in the TC limit:
for every quanta $\theta_q$ absorbed from $T_q$, a quanta $\theta_w$ must be absorbed from the work bath. The efficiency and $\eta^{(n)}$ therefore  obey the following relations \cite{Segal18},
\bea
\eta
&\equiv&\frac{\langle q\rangle}{\langle w \rangle}
=\frac{\theta_q}{\theta_w}, \,\,\,\,\,\
\nonumber\\
\eta^{(n)} &\equiv&\frac{\langle \langle q^n\rangle\rangle}{\langle \langle w^n \rangle\rangle} = 
\left(\frac{\theta_q}{\theta_w}\right)^n  = \eta^n \,\,\,\,\, n=2,3,...
\label{eq:def12}
\eea
Furthermore, the cooling condition is \cite{Segal18}
\bea
\langle q\rangle &\propto& n_q(\theta_q) n_w(\theta_w) [n_h(\theta_h)+1]
\nonumber\\
&-& 
[ n_q(\theta_q)+1][ n_w(\theta_w)+1] n_h(\theta_h) \geq 0, 
\eea
with $n_{i}(\theta_{i}) = \frac{1}{e^{\beta_{i}\theta_i}-1}$ the Bose Einstein distribution function of the bath $i=h,w,q$ with
transition $\theta_i$. The cooling condition can be equivalently written as
\bea
\frac{\theta_q}{\theta_w} \leq \frac{\beta_h-\beta_w}{\beta_q-\beta_h},
\eea
with $\beta_i=1/T_i$ the inverse temperature.
We identify the left hand side by the efficiency, $\eta$, and the right hand side by the Carnot efficiency, thus
\bea
\eta\leq \eta_C = \frac{\beta_h-\beta_w}{\beta_q-\beta_h}.
\eea
%
Altogether, an individual 3LQAR operates in the TC limit and it satisfies
 \bea
 \eta^n= \eta^{(n)}\leq \eta_C^n.
 \label{eq:eta2S}
 \eea
 
\subsection{Ensemble of $N>1$ distinct refrigerators}
An ensemble of distinct uncorrelated refrigerators, with possibly different spacings $\theta_q$ and $\theta_w$ and different system-bath coupling energies, provides a nontrivial setting for exemplifying the lower bound on ratios of fluctuations. We represent such an ensemble in Fig. \ref{fig:QAR}(c).
We assume that all our systems are operating between the same heat baths, $T_i$; $i=w,h,q$, and we fix $\theta_h$.

First, given the validity of the upper bound, Eq. (\ref{eq:eta2S}) for each individual machine, we conclude that a similar upper bound holds for the ensemble, as we proved in Sec. \ref{SecU}. In what follows we therefore focus on establishing a lower bound on $\eta_N^{(n)}$.


\begin{figure*} \hspace{-16mm}
{\hspace{-6mm}\includegraphics[width=20cm]{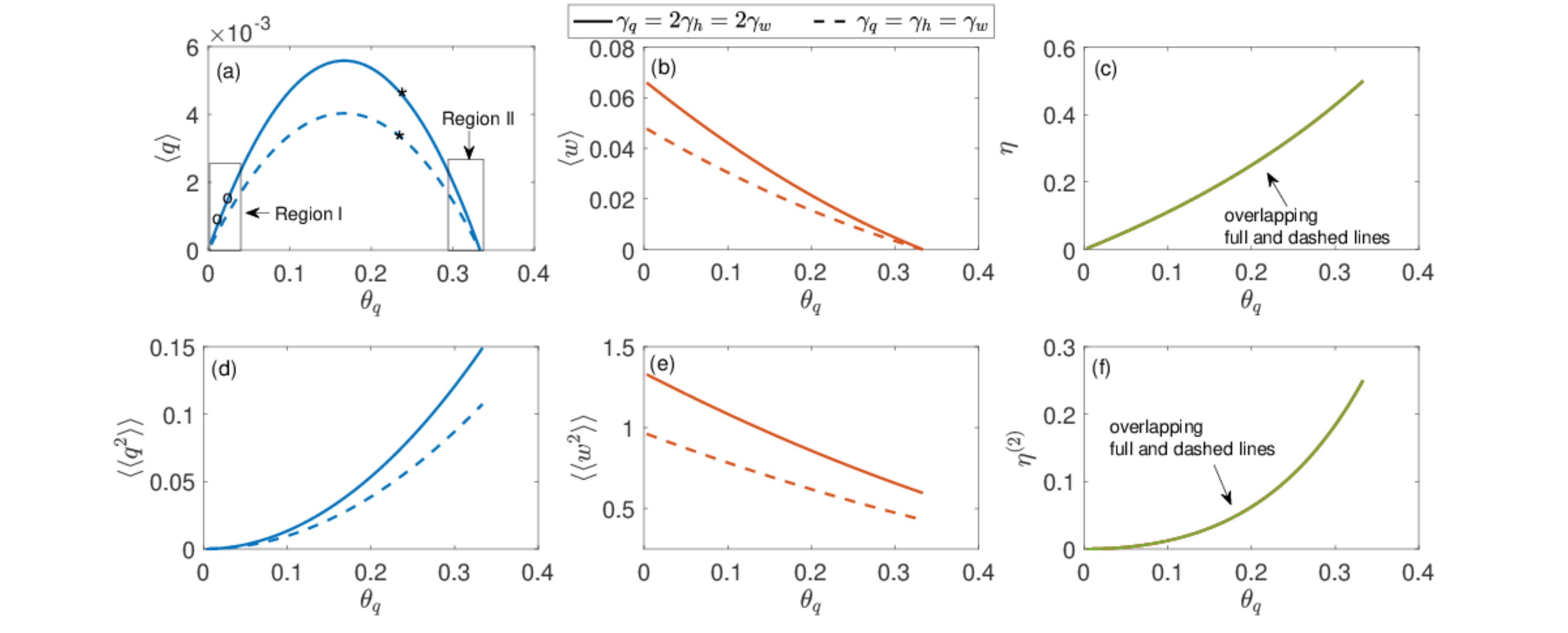} 
\caption{Exemplifying the performance of 3LQARs operating between $\beta_q$=0.4, $\beta_h$=0.2,  and $ \beta_w$=0.1.
(a) Cooling current, (b) current extracted from the work bath,
(c) cooling efficiency, (d) fluctuations of cooling current, (e) fluctuations of the so-called work current, and (f) the relative noise $\eta^{(2)}$.
Other parameters are  $\theta_h=1$, and system-bath couplings
 $\gamma_q=\gamma_h=\gamma_w$ (dashed), 
$\gamma_q=2\gamma_h=2\gamma_w$ (full).
In Sec. \ref{sec:QARgaps},
We prove the lower bound $\eta_N^{n} \leq \eta_N^{(n)}$ in Region I. In Appendix A, it is proved to hold in Region II, albeit only for $n=2$ and $N=2$.
}}
\label{fig:curr}
\end{figure*}
%


\subsubsection{QARs with different system-bath couplings}
\label{QAR-as}

We consider an ensemble of three-level systems, each with different coupling strength to the baths but with the same gaps $\theta_q$  and  $\theta_w$. Members of this ensemble sit vertically (at the same $\theta_q$)
 along different curves as exemplified in Fig. \ref{fig:curr}(a) with asterisks. 
While the heat currents depend on both the coupling parameters and the energy spacings, notably for any individual refrigerator the ratio $\eta^{(n)}$ depends only on the latter.
Next we prove the saturation of the lower bound,
\bea
\eta_N^n = \eta_N^{(n)}.
\label{eq:low1}
\eea
%
For each 3LQAR, $\langle w_k\rangle/\langle q_k\rangle  =\theta_w/\theta_q \equiv \alpha$, see Eq. (\ref{eq:def12}).
Therefore, the efficiency of the ensemble of refrigerators to the power $n$ is
\bea
\eta_N^n &=& \left(\frac{\sum_{k=1}^N \langle q_k\rangle }{\sum_{k=1}^N \langle w_k\rangle}\right)^n 
\nonumber\\
&=&\left(\sum_j \frac{\langle q_j\rangle}{\langle w_j\rangle} \frac{\langle\omega_j\rangle}{\sum_k \langle w_k \rangle} \right)^n
= \frac{1}{\alpha^n}.
\eea
Similarly, based on Eq. (\ref{eq:def12}),
\bea
\eta_N^{(n)} = \left(\frac{\sum_{k=1}^N \langled q_k^n\rangled }{\sum_{k=1}^N \langled w_k^n\rangled}\right)^n = \frac{1}{\alpha^n},
\eea
and we confirm Eq. (\ref{eq:low1}).

\begin{figure*}[htbp]
{\includegraphics[width=15.0cm] {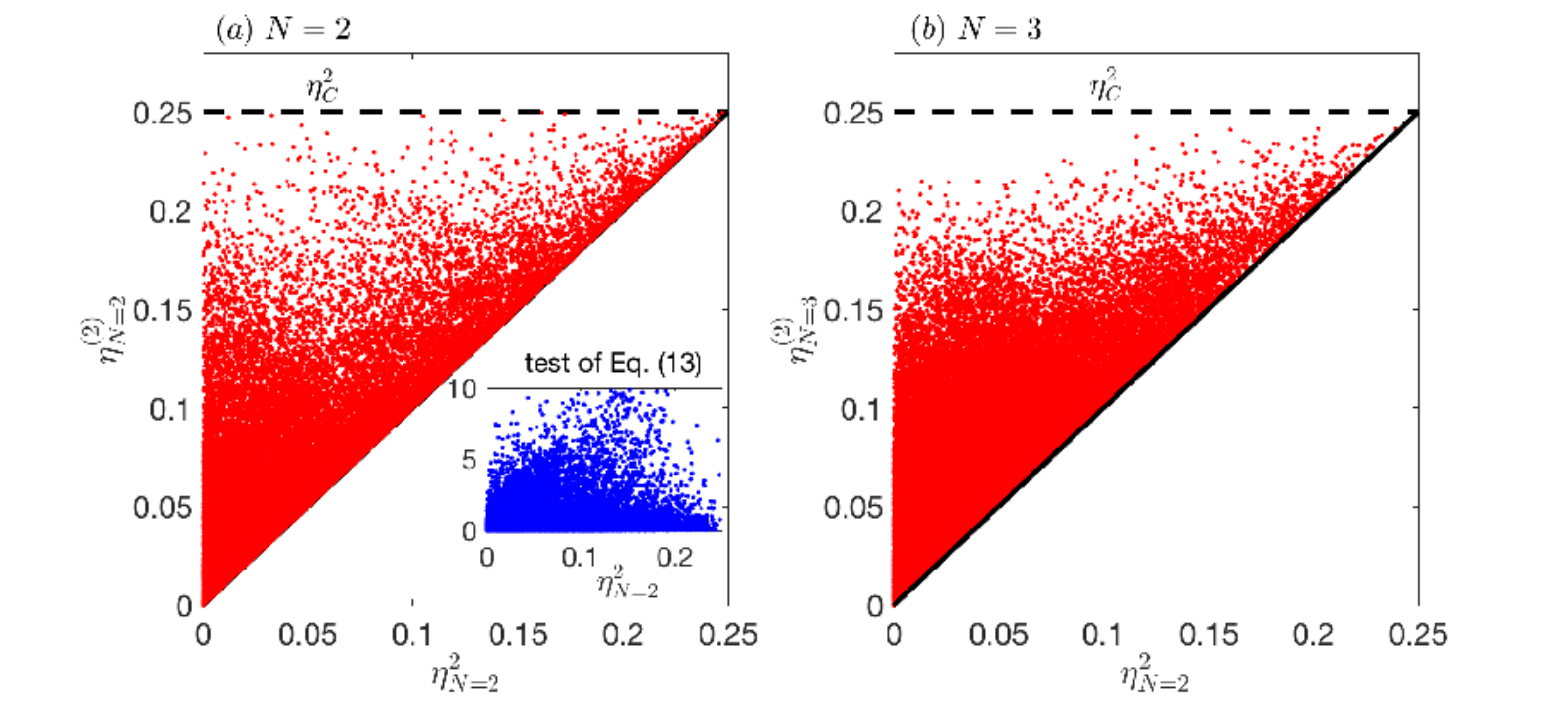} 
\caption{Demonstration that the lower bound $\eta_N^2\leq \eta^{(2)}_N$ is valid beyond Regions I and II of the cooling window.
(a) Each dot corresponds to a pair of three-level systems
with random values for $\theta_q$ and  system-bath couplings, $\gamma_i$. 
For this pair, we calculate the total cooling and work currents, as well as their fluctuations. 
In the inset we test the inequality  (\ref{eq:low3}), which is stronger than the lower bound: 
We present the difference $\langled w_A^2\rangled/ \langled w_B^2\rangled-
\langle w_A\rangle/ \langle w_B\rangle$ for pairs with $\eta_A\geq \eta_B$ and show that it is always positive.
(b) Same calculations as in (a), but for a triplet of 3LQARs.
Temperatures are the same as in Fig. \ref{fig:curr}.
Coupling strengths are uniformly sampled from $0\leq\gamma_i\leq 1$, 
$\theta_q$ is sampled from the cooling region.
  }}
\label{fig:boundN}
\end{figure*}
%

\subsubsection{QARs with distinct gaps}
\label{sec:QARgaps}

We now prove the lower bound for an ensemble of 3LQARs characterized by {\it distinct} energy gaps, but coupled in the same manner to the different baths; we highlight that each three-level system may couple asymmetrically to the three different baths, but all the three-level systems follow the same coupling parameters. 
The proof described in this Section holds in the low-cooling regime identified as Region I in Fig. \ref{fig:curr}(a);
points marked by circles exemplify members of this ensemble.
In Appendix A we describe a complementary proof (for $N=2$ and $n$=2) that holds in Region II,  at the edge of the cooling window marked
in Fig. \ref{fig:curr}(a). 

Henceforth, for simplicity and without loss of generality we set the total gap at $\theta_h=1$.
Assuming that $\theta_q\ll1$, we Taylor-expand the cooling current,
\bea
\langle q \rangle \approx \alpha \theta_q + \beta \theta_q ^2 +...
\label{eq:qd}
\eea
Plugging this expansion into 
$\eta=\frac{\langle q \rangle }{\langle w\rangle } = \frac{\theta_q}{\theta_w}$ provides a consistent expansion for the input work,
\bea
\langle w \rangle \approx \alpha -(\alpha-\beta)\theta_q + ...
\eea
%
%
Similarly, we write a Taylor expansion for the fluctuations of the cooling current,
\bea
\langled q^2 \rangled\approx  \gamma \theta_q^2 + \delta \theta_q^3+...
\eea 
and using $\eta^{(2)} = \frac{\theta_q^2}{\theta_w^2}=\frac{\langled q^2 \rangled }{\langled w^2\rangled }$
 put together  the consistent expansion for the fluctuations of the  work current,
\bea
\langled w^2 \rangled\approx  \gamma +  \theta_q(\delta-2\gamma) + ...
\eea
Since we assume small cooling current 
and correspondingly small fluctuations (see Fig. \ref{fig:curr}), $\theta_q\ll1$, $\beta\theta_q \ll \alpha$,  and $ \delta\theta_q\ll \gamma$, the lowest order expansions for the currents and their fluctuations are,
 \bea
 \langle q\rangle &\approx& \alpha \theta_q, \,\,\,\,\ \langle w\rangle\approx \alpha.
 \nonumber\\
 \langled q^2 \rangled &\approx& \gamma \theta_q^2, \,\,\,\,
\langled w^2 \rangled \approx \gamma.
\label{eq:qw2d}
 \eea
We are now ready to test the lower bound. We begin with two refrigerators A and B 
of spacings $\theta_{qA}$ and $\theta_{qB}$
that are coupled in the same manner to the baths (for example, A and B are marked by circles in Fig. \ref{fig:curr}). Since these systems
lie on the same curve, 
 $\langle q_A\rangle \approx \alpha \theta_{qA}$, $\langle q_B\rangle \approx \alpha \theta_{qB}$,
  $\langle w_A\rangle \approx \langle w_A\rangle\approx\alpha$. 
  Similarly, 
 $\langled q_A^2\rangled \approx \gamma \theta_{qA}^2$, $\langled q_B^2\rangled \approx \gamma \theta_{qB}^2$,
  $\langled w_A^2\rangled \approx \langled w_B^2\rangled\approx\gamma$.
We now test the inequality
\bea
\eta^2_{N=2}=\left( \frac{ \langle q_A\rangle +  \langle q_B\rangle}{\langle w_A\rangle + \langle w_B\rangle}\right)^2
\leq \frac{  \langled q_A^2\rangled +  \langled q_B^2\rangled }{\langled w_A^2\rangled + \langled w_B^2\rangled} = \eta^{(2)}_{N=2},
\nonumber\\
\label{eq:lowerd}
\eea
by substituting the currents and fluctuations. It reduces to
\bea
\frac{(\theta_{qA} + \theta_{qB})^2}{4} \leq \frac{\theta_{qA}^2 + \theta_{qB}^2}{2},
\eea
which is true since $(\theta_{qA}-\theta_{qB})^2\geq 0$. 

This proof can be generalized to the bound on $\eta^{(n)}$ for an ensemble of $N$ refrigerators. Given the small-$\theta_q$ expansions of the $n$th cumulant \cite{Segal18}, $\langled q^n_k \rangled \approx \gamma_n \theta_{qk}^n$, $ \langled w^n_k \rangled \approx \gamma_n$, the lower-bound inequality reads

\bea \eta^n_N = \left(\frac{\sum_k \theta_{qk}}{N}\right)^n \leq 
\frac{\sum_k \theta_{qk}^n}{N} = \eta^{(n)}_N.
\eea
Taking the $n$th root gives the desired result by the power mean inequality\cite{ineqHandbook}, with power 1 on the left hand side and $n$ on the right,
\bea
\eta_N=\left(\frac 1N \sum_k \theta_{qk}\right) \leq 
\left(\frac 1N \sum_k \theta_{qk}^n\right)^{1/n}=\left[\eta_N^{(n)}\right]^{1/n}.
\nonumber\\
\label{eq:PMI}
\eea
The power mean inequality also implies that $n = 2$ provides the tightest bound on $\eta_N$.

We note that in general, the assumption of the cooling current being linear in the gap, 
$\langle q\rangle = \alpha \theta_q$, does not necessarily correspond to a linear response limit, $\langle q\rangle\propto (T_w-T_q)$. However, for the 3LQARs, this is in fact the case \cite{Segal18}:
A linear dependence of $\langle q\rangle$ with $\theta_q$ develops hand in hand with the current becoming linear in the temperature difference, 
though the temperature difference could be large, $\Delta T/T_i\gg 1 $.

In Appendix A, we prove the lower bound in Region II as marked in Fig. \ref{fig:curr}, albeit limited to  $n=2$ and $N=2$.
\subsubsection{Simulations}
\label{sec:simulQAR}
We exemplify in Fig. \ref{fig:curr} the behavior of individual 3LQARs. The population dynamics, heat currents and their fluctuations are calculated as described in Ref. \cite{Segal18} using kinetic-like quantum master equations. The three-level system is coupled to the heat baths with an Ohmic spectral density function, with the excitation rate constant e.g. between the lowest state to the intermediate one, $k_{q}= \Gamma_q(\theta_q) n_i(\theta_q)$ where we use an Ohmic model,  $\Gamma_q(\theta_q)=\gamma_q\theta_q$; $n_i(\theta_i)$ is the Bose-Einstein distribution function. 
The relaxation rate constant follows from local detailed balance. 
$\gamma_i$ are dimensionless coefficients that control the coupling strength.
Simulations in  Fig. \ref{fig:curr} agree with theoretical results for the efficiency 
$\eta= \theta_q/\theta_w$ and the ratio of fluctuations, 
$\eta^{(2)}= (\theta_q/\theta_w)^2$.

In Fig. \ref{fig:boundN}(a), we present simulation results for $\eta^{(2)}_N$. 
We generate a randomized set of 300 refrigerators by sampling uniformly 
over different values of $\theta_q$ within the cooling domain. As for the coupling strength,
they are sampled from a uniform distribution $0\leq\gamma_i\leq 1$, and they can be distinct and unequal at each realization  ($\gamma_h\neq\gamma_q\neq\gamma_w$).
We then consider every possible pair of refrigerators and calculate for this pair
the total cooling ($\langle q_A\rangle + \langle q_B\rangle$) and work currents ($\langle w_A\rangle + \langle w_B\rangle$), as well as their fluctuations,  $\langled q_A^2\rangled + \langled q_B^2\rangled$ and $\langled w_A^2\rangled + \langled w_B^2\rangled$.
This allows us to calculate the efficiency for each pair, $\eta_{N=2} = (\langle q_A\rangle +\langle q_B\rangle)/(\langle w_A\rangle+\langle w_B\rangle)$, and the ratio of their fluctuations, $\eta_{N=2}^{(2)} $. 
Though most of our refrigerators lie outside Regions I and II,
we do not observe violations to the lower bound, $\eta_{N=2}^2 \leq \eta_{N=2}^{(2)}$; note that
 parameters are outside the linear-response regime \cite{Gerry}.

In Fig. \ref{fig:boundN}(b), we select 90,000 samples of triplet 3LQARs and calculate their total current and fluctuations;
the total cooling current is  $\langle q_A\rangle + \langle q_B\rangle+\langle q_C\rangle$ and the associated fluctuation is
  $\langled q_A^2\rangled + \langled q_B^2\rangled +\langled q_C^2\rangled$. We then obtain the efficiency for a triplet 3LQAR
  and the ratio over fluctuations of these machines.
Again we confirm based on simulations that the lower bound (\ref{eq:low}) is satisfied.

\subsubsection{Discussion}
We organize our observations up to this point on the validity of the bounds (\ref{eq:goal}) for an ensemble of tight-coupling refrigerators. 
First, the upper bound holds without additional assumptions (Sec. \ref{SecU}).
As for a lower bound on $\eta^{(n)}$, 
for the 3LQARs as discussed in this Sec., we proved that:
(ii) The lower bound saturates for an ensemble of systems with identical spacings, $\theta_q$, but distinct couplings to the baths.
(iii) The lower bound holds for an ensemble of systems operating with equal coupling schemes. This proof hold in the limit of small cooling currents, in Region I, characterized by a vanishing $\theta_q$. We were also able to prove (Appendix A) the lower bound in Region II, characterized by a vanishing input work current, albeit  limited to  $n=2$ and $N=2$.
(iv) Extensive simulations confirmed the validity of the lower bound more broadly beyond linear response. 
Even more so, our simulations showed that an
inequality more general than the lower bound holds, namely Eq. (\ref{eq:low3}).
(v) In Appendix B we discuss the validity of the lower bound for a three-level model operating as an engine.


\section{Model II: Ensemble of tight-coupling thermoelectric junctions}
\label{SecTE}

\begin{figure}[b]
{\includegraphics[width=0.9\columnwidth] {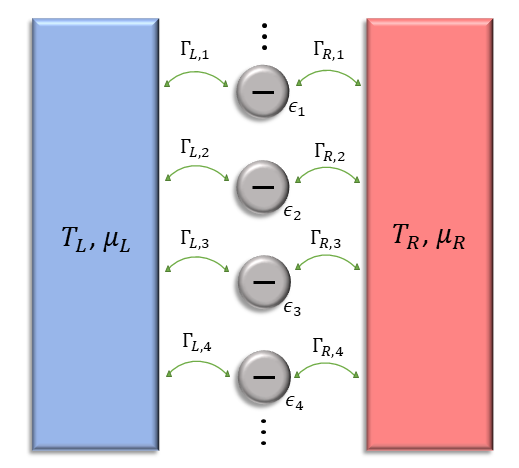} 
\caption{
An ensemble of thermoelectric junctions,  
made of a collection of noninteracting and uncorrelated quantum dots each characterized by a resonant level $\epsilon_k$ and coupling strengths $\Gamma_{L,k}$ and $\Gamma_{R,k}$ to the left and right lead, respectively. The junctions all operate between the same metal leads.
}}
\label{fig:te_illustration}
\end{figure}

A thermoelectric junction comprises a left $(L)$ and right $(R)$ metal lead, between which charge and energy transport may be mediated through an embedded quantum system. The leads serve as thermal baths, differing in temperature and chemical potential such that the associated affinities oppose one another\cite{benenti17}; we  suppose that $T_L < T_R$ and $\mu_L>\mu_R$. In what follows, we assume resonant electron transport through a discrete level \cite{LinkeE}, and discuss the existence of the lower bound on $\eta^{(n)}_N$ for an ensemble with $N$ thermoelectric junctions as depicted in Fig. \ref{fig:te_illustration}.

\subsection{Single junction, $N=1$}
We identify a stochastic charge current $j_k$, for a junction labelled by $k$, focusing here on the tight-coupling limit \cite{Caplan,benenti17}, wherein the mean energy current is $\epsilon_k\langle j_k\rangle$--proportional to the charge current, where $\epsilon_k$ is the resonant level of a single quantum dot characterizing the system through which transport is mediated. In this limit, the charge current and fluctuations are given by 
\cite{Levitov,schonhammer07}
\bea\label{eq:te_general}
    \langle j_k\rangle &=& \mathcal{T}_1\left[f_L(\epsilon_k) - f_R(\epsilon_k)\right],
    \nonumber\\
    \langled j_k^2\rangled &=& \mathcal{T}_1\left[f_L(\epsilon_k)(1 - f_R(\epsilon_k)) + f_R(\epsilon_k)(1-f_L(\epsilon_k)\right]
    \nonumber\\
    & &-\mathcal{T}_2\left[f_L(\epsilon_k) - f_R(\epsilon_k)\right]^2.
\eea
The coefficients $\mathcal{T}_1$ and $\mathcal{T}_2$ are determined by the coupling strengths $\Gamma_L$ and $\Gamma_R$ between the system and the two leads, 
$\mathcal{T}_1=\frac{\Gamma_L\Gamma_R}{\Gamma_L+\Gamma_R}$ and $\mathcal {T}_2=\frac{2\Gamma_L^2\Gamma_R^2}{(\Gamma_L+\Gamma_R)^3}$ for the resonant-level model, see Refs. \cite{Bijay-TUR,Junjie-TUR}.
$f_L(\epsilon)$ and $f_R(\epsilon)$ are the Fermi-Dirac distributions for the two leads. A thermoelectric device may act as a refrigerator or an engine in various operational regimes, as determined by the directions of power and heat currents, which are proportional to the charge current in the TC limit.
The mean currents thus obey $\langle w_k\rangle = \Delta\mu\langle j_k\rangle$, $\langle q_k\rangle = (\epsilon_k - \mu_\nu)\langle j_k\rangle$ ($\nu = L$ for a refrigerator, $R$ for an engine)
with $\Delta \mu=\mu_L-\mu_R$.
The fluctuations in these quantities are similarly determined by those for the charge current: $\langled w_k^2\rangled = \Delta\mu^2\langled j_k^2\rangled$, $\langled q_k^2\rangled = (\epsilon_k - \mu_\nu)^2\langled j_k^2\rangled$. 

An individual tight-coupling thermoelectric junction has been shown to satisfy the equality $\eta_k^n = \eta_k^{(n)}$ for any $n = 1,2,3,...$\cite{Gerry}, where, with $\epsilon_k$ the dot energy for the machine $k$,
\bea\label{eq:tc_eta}
    \eta_k &=& \frac{\Delta\mu}{\epsilon_k - \mu_R}\:\textrm{(engine)}
    \nonumber\\
    \eta_k &=& \frac{\epsilon_k - \mu_L}{\Delta\mu}\:\textrm{(refrigerator)}.
\eea
Immediately, the upper bound, $\eta_k^{(n)}\leq\eta_C^n$, also holds\cite{Gerry} where $\eta_C$ represents the operational Carnot bound.

\begin{figure*}[t]
{\includegraphics[width=14.0cm, trim=60 25 60 25] {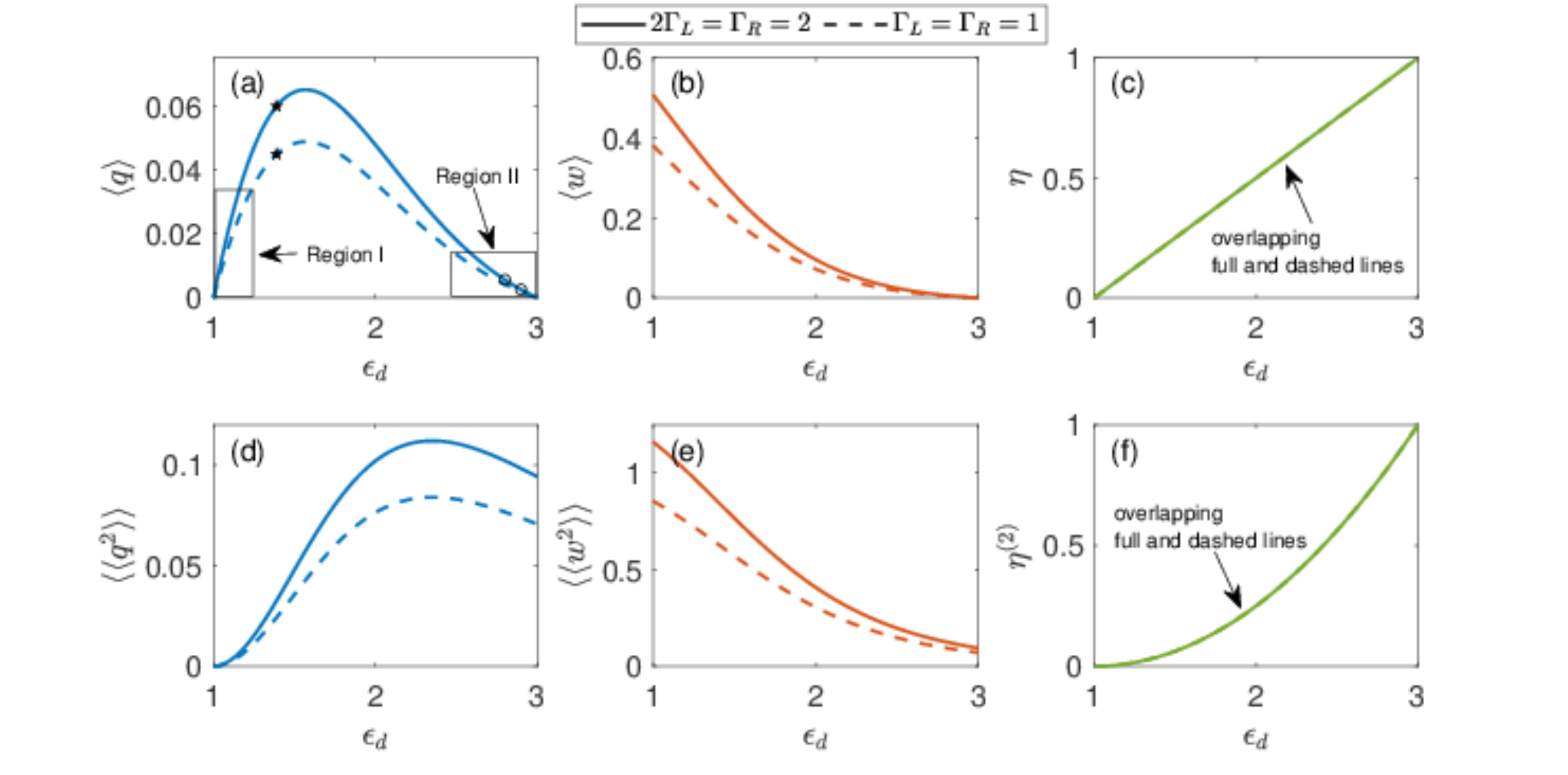}
\caption{Exemplifying the performance of individual thermoelectric refrigerators between $\beta_L$=2, $\beta_L$=1, $\mu_L = 1$, and $ \mu_R=-1$. (a) Cooling current, (b) work current, (c) cooling efficiency, (d) fluctuations of cooling current and (e) work current, and (f) the relative noise $\eta^{(2)}$.
In Sec. \ref{Sec:TED} we prove the lower bound in Region II under some conditions.
Region I is explored in Appendix C.
}}
\label{fig:single_ref}
\end{figure*}
%

\subsection{Ensemble of distinct thermoelectric junctions}

We now consider an ensemble of distinct tight-coupling thermoelectric junctions, labelled by $k$, whose internal parameters such as $\epsilon_k$ and system-lead coupling strengths may differ, but which operate between leads with the same set of bath parameters. 
For example, we envision a thermoelectric device 
with parallel quantum dots each conducting resonantly with their own energy and coupling strengths to the metal electrode, see Fig. \ref{fig:te_illustration}. 

The validity of the upper bound for a single junction, along with the general result expressed in Eq.~(\ref{eq:etaNn}), leads to the analogous upper bound  for an ensemble of such junctions as discussed in Sec. \ref{SecU}. The following discussion will therefore focus on when the lower bound, $\eta^2\leq\eta^{(n)}$, holds for such an ensemble.

\subsubsection{Thermoelectric junctions with different system-bath couplings}
\label{Sec-TEd}

Consider an ensemble of $N$ thermoelectric junctions, labelled by $k$, with the same resonant level, $\epsilon$, characterizing the quantum system, as represented in Fig.~\ref{fig:single_ref}(a) by the pair of asterisks. 
The cooling efficiency, $\langle q_k\rangle/\langle w_k\rangle = (\epsilon - \mu_L)/\Delta\mu$ is equal for all such refrigerators since it does not depend on the system-bath coupling.
As such, one can readily prove, via an argument mirroring that in Section \ref{QAR-as}, the strict equality $\eta_N^n=\eta_N^{(n)}$ for the ensemble.

\begin{figure*}[t]
\includegraphics[width=15.0cm] {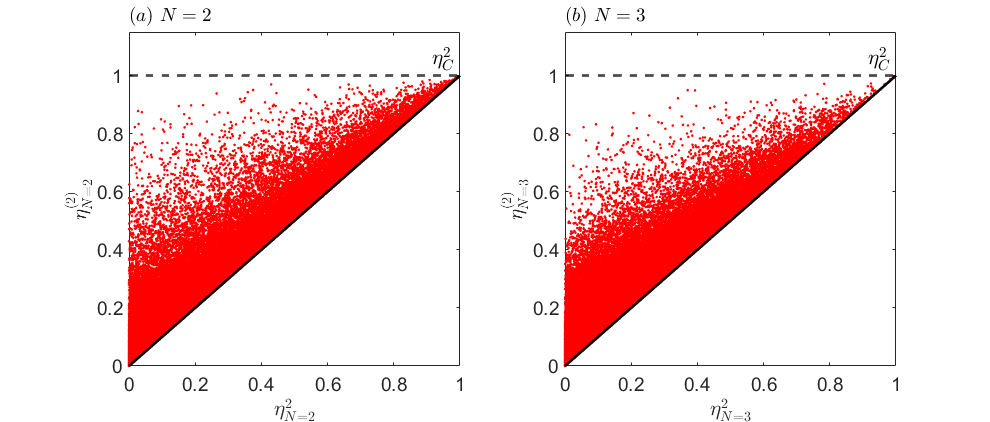} 
\caption{Demonstration that the lower bound $\eta_N^2\leq \eta^{(2)}_N$ is valid beyond the linear regime of the charge current.
(a) Each dot corresponds to a pair of tight-coupling thermoelectric refrigerators
with random values for $\epsilon_k$ as well as coupling strengths. For this pair, we calculate the total cooling and work currents as well as their fluctuations. (b) Same calculations as in (a), but for a triplet of thermoelectrics.
$\beta_L = 2$, $\beta_R = 1$, $\mu_L = 1$, and $\mu_R = -1$. Coupling strengths are uniformly sampled from $0\leq\Gamma_i\leq1/2$, 
$\epsilon_k$ is sampled from the cooling region, $1<\epsilon_k<3$. The operational Carnot bound for refrigeration is $\eta_C = \beta_R/(\beta_L - \beta_R)$.
}
\label{fig:boundN_thermoelectric}
\end{figure*}


\subsubsection{Thermoelectric junctions with distinct energy levels}
\label{Sec:TED}
Now, we will focus specifically on the case of a pair ($N=2$) of tight-coupling thermoelectric junctions, $A$ and $B$, operating as refrigerators, with cooling currents flowing from the left (cold) lead, $\langle q_k\rangle = (\epsilon_k - \mu_L)\langle j_k\rangle$, $k=A,B$. The cooling and work currents are taken, by convention, to be positive. We will show that for this pair taken as a single device, the lower bound, $\eta_{N=2}^2\leq\eta_{N=2}^{(2)}$ holds, provided we restrict ourselves to a specific range of values for $\epsilon_A$ and $\epsilon_B$, namely, Region II, near the edge of the cooling window such that $\langle j_k\rangle$ is small. For instance, such an ensemble may be represented by the pair of circles in Fig.~\ref{fig:single_ref}(a). Furthermore, we will suppose that one refrigerator, say $B$, has a significantly larger charge current than the other ($\langle j_A\rangle/\langle j_B\rangle\ll1$). This requires that $\epsilon_A>\epsilon_B$, so $\eta_A>\eta_B$. 
The complementary proof for Region I is given in Appendix C.

We express the dot energy for each thermoelectric refrigerator as $\epsilon_k = \mu_R + \Delta\mu\beta_L/\Delta\beta - u_k$
with $\Delta \beta=\beta_L-\beta_R$, $\Delta \mu=\mu_L-\mu_R$.
Note that the charge current vanishes when $u_k=0$ \cite{Linke05}, and initially grows linearly with $u_k$: $\langle j_k\rangle \approx \alpha u_k$, where $\alpha$ is some constant coefficient. Then we have that $u_A/u_B\ll1$, and we may write out the full efficiency of the pair, 
$\eta_{N=2}=[(\epsilon_A-\mu_L)u_A + (\epsilon_B-\mu_L)u_B]/[(u_A+u_B)\Delta \mu]$.
Truncating after first order in this ratio, and squaring, we get,
\be 
    \eta_{N=2}^2\approx\eta_B^{(2)}\left[1 + 2\bigg(\frac{\epsilon_A - \mu_L}{\epsilon_B - \mu_L} - 1\bigg)\frac{u_A}{u_B}\right].
\ee
The assumption of small charge current permits the approximation that fluctuations are predominantly due to 
single-electron processeses;
we refer to Eq.~(\ref{eq:te_general}) and ignore the contribution proportional to $(f_L(\epsilon_k) - f_R(\epsilon_k))^2$, writing
\be 
    \langled j_k^2\rangled = \mathcal{T}_1\left[f_L(\epsilon_k)(1 - f_R(\epsilon_k)) + f_R(\epsilon_k)(1-f_L(\epsilon_k))\right].
\ee
Thus can be shown to be proportional to the charge current,
\be 
    \langled j_k^2\rangled = \langle j_k\rangle \coth\left(\frac{\Delta\beta u_k}{2}\right)\equiv
    c_k\langle j_k\rangle,
\ee
with $c_k$ defined from this relation.
The ratio of fluctuations of the cooling current to work current is given in terms of $u_A$ and $u_B$ as
\be
    \eta_{N=2}^{(2)} = \eta_B^{(2)}\frac{1 + \frac{(\epsilon_A - \mu_L)^2}{(\epsilon_B - \mu_L)^2}\frac{c_Au_A}{c_Bu_B}}{1 + \frac{c_Au_A}{c_Bu_B}}.
\ee
As $u_k$ approaches zero, $c_k$ goes to infinity, so we cannot assume $c_Au_A/c_Bu_B$ is small. However, we can still compare $\eta_{N=2}^{2}$ and $\eta_{N=2}^{(2)}$, finding, after keeping only terms up to first order in $u_A/u_B$, that the lower bound is equivalent to the inequality
\be
    2\bigg(\frac{\epsilon_A - \mu_L}{\epsilon_B - \mu_L} - 1\bigg)\leq \frac{c_A}{c_B}\left[\frac{(\epsilon_A - \mu_L)^2}{(\epsilon_B - \mu_L)^2} - 1\right].
\ee
Since $\epsilon_A>\epsilon_B$, this holds as long as $c_A/c_B\geq 2$. We note that this is certainly the case since, by assumption, $u_A$ and $u_B$ are small, so $c_A/c_B\approx u_B/u_A\gg1$. We may conclude that in this regime, $\eta_{N=2}^2\leq\eta_{N=2}^{(2)}$. 

The proof outlined in this section is quite limited: It holds for a pair of refrigerators and only for the second cumulant, $n=2$.
Complementing this proof, in Appendix C, we prove the lower bound in Region I, now for an ensemble of arbitrary size $N$ and to the order $n$. Furthermore, Fig.~\ref{fig:boundN_thermoelectric} demonstrates, via the results of simulations, that the upper and lower bounds (for $n=2$) appear to hold for pairs and triplets of thermoelectric junctions outside Regions I and II.

Altogether, we proved the validity of the lower bound for three-level refrigerators and thermoelectric refrigerators in corresponding cases, in Region I and II.
In Appendix D, we consider an analogous set of assumptions as discussed in this subsection, but applied to thermoelectric junctions operating as {\it engines}. We find in this case that the lower bound must be violated.

\bigskip


\section{Summary}
We studied universal bounds on ratios of fluctuations, as well as higher order cumulants, in an ensemble of distinct, uncorrelated thermal machines operating in steady state and arbitrarily far from equilibrium.
We built our bounds for the ensemble of machines from bounds on the individual machine, $\eta_k^n=\eta_k^{(n)}\leq\eta_C^n$. These relations hold e.g., for systems obeying the tight coupling limit, such as three-level absorption refrigerators and resonant-level thermoelectric junctions.  $\eta_C$ is the Carnot efficiency and $\eta^{(n)}$ is ratio of the $n$th cumulants of the output current to input current.
We proved that:

(i) An upper bound holds for an ensemble of $N$
distinct machines, engines or refrigerators, arbitrarily far from equilibrium, $\eta_N^{(n)}\leq \eta_C^n$.


 (ii) The lower bound on the ensemble, $\eta_N^n\leq\eta_N^{(n)}$, is more limited in scope. While it seems to hold  for refrigerators, based on analytic and numerical work, 
  we demonstrated that thermoelectric {\it engines} may violate the lower bound for $N\geq2$.
  
(iii) Focusing on three-level quantum absorption refrigerators and thermoelectric junctions, we proved the validity of the lower bound,   $\eta_N^n\leq\eta_N^{(n)}$, in different cases: when the cooling current was small, and for an ensemble of distinct machines with identical efficiencies. 
  
  (iv)  While analytic proofs of the lower bound for refrigerators were limited in scope (e.g. to $n=2$ and $N=2$ in Region II near the edge of the cooling window), 
 simulations beyond these strict regimes support its validity more broadly.
 
The significance of our study is in developing bounds for ensemble of machines. While bounds on the individual system may be trivial, as in the tight coupling limit, a machine that collects input and output from {\it several} tightly-coupled components (recall Figs. \ref{fig:QAR} and \ref{fig:te_illustration}) may behave nontrivially in this respect, even under classical laws and in the absence of interactions between components.
Adding interactions to the working fluid e.g. Coulomb interactions between quantum dots in a themroelectic device or
by using an interacting atomic gas, opens up the door to new effects, such as many-particle enhancement of performance \cite{delCampoEnt16,delCampoNJP16,delCampoNPJ19,Reimann,Gernot19,Gernot}. Furthermore,  quantum statistical effects combined with interactions or collective unitary operations on the components can further enhance performance, as predicted in Ref. \cite{delCampo}.

An important result of our work is in tightening bounds on efficiency: As an outcome of the lower bound, Eq. (\ref{eq:PMI}), one immediately finds that $\eta\leq \left[\eta^{(n)}\right]^{1/n}\leq \eta_C$,
and that the case with $n=2$ provides the tightest lower bound. Whether this is a general result, or only valid for refrigerators in the small-$\theta$ small-cooling domain
remains an open question. 

In future work we will focus on understanding the validity of the lower bound on $\eta^{(n)}$ in general settings, and on exploring the fundamental differences between engines and refrigerators as pertain to bounds on fluctuations. 




\begin{acknowledgements}
DS acknowledges support from an NSERC Discovery Grant and the Canada Research Chair program.
The work of NK was supported by the Ontario Graduate Scholarship (OGS). The research of MG was supported by the NSERC Canada Graduate Scholarship-Master's and the OGS. 

NK and MG contributed equally to this work and are joint ``first authors".
\end{acknowledgements}

\begin{widetext}
\section*{Appendix A: Lower bound on $\eta_2^{(2)}$ for 3LQARs at the boundary of the cooling region (Region II)}
\setcounter{equation}{0}  
\renewcommand{\theequation}{B\arabic{equation}}
\setcounter{equation}{0}  
\renewcommand{\theequation}{A\arabic{equation}}
\label{app:1}


Here, we present a proof of the validity of a lower bound for two 3LQARs, applicable in the limit of small currents.
This proof holds at the boundary of the cooling domain (see Region II in Fig. \ref{fig:curr}).
The proof presented here is limited to two QARs, and its extensions to $N$ refrigerators is not obvious.

For simplicity, we define the scaled currents $q/\theta_q\to q$ and  $w/\theta_w\to w$. We also assume that $T_w\to \infty$.
The cooling current and its noise are given in terms of the bath-induced transition rate constants $\Gamma_{q,h,w}$, see text in Sec. \ref{sec:simulQAR} and Ref. \cite{Segal18}.
We also use the short notation $\tilde\Gamma_{i}=\Gamma_{i}(\theta_i)n_i(\theta_i)$.
The cooling current and its fluctuations are given by \cite{Segal18}
\bea
 \langle q \rangle = \frac{\tilde \Gamma_h\tilde\Gamma_q\tilde\Gamma_w\left(e^{\theta_h/T_h}-e^{\theta_q/T_q}\right)}{M}, 
 \eea
\bea \nonumber 
\langled q^2 \rangled = \frac{2\left(\tilde\Gamma_h+\tilde\Gamma_q+\tilde\Gamma_w+\tilde\Gamma_qe^{\theta_q/T_q}+
\tilde\Gamma_w+\tilde\Gamma_he^{\theta_h/T_h}\right)\langle q\rangle^2}{M} + 
\frac{\tilde\Gamma_q\tilde\Gamma_h\tilde\Gamma_w\left(e^{\theta_h/T_h}+e^{\theta_q/T_q}\right)}{M}
\eea
\bea \nonumber
\text{with } M = \left(\tilde\Gamma_ce^{\theta_q/T_q} + \tilde\Gamma_w\right)\left(\tilde\Gamma_w + \tilde\Gamma_qe^{\theta_h/T_h}\right) + \left(\tilde\Gamma_q+\tilde\Gamma_h\right)\left(\tilde\Gamma_w+\tilde\Gamma_he^{\theta_q/T_q}\right) +\nonumber \\ \left(\tilde\Gamma_q+\tilde\Gamma_h\right)\left(\tilde\Gamma_qe^{\theta_q/T_q} + \tilde\Gamma_w\right) - \tilde\Gamma_q^2e^{\theta_q/T_q} - \tilde\Gamma_w^2 -\tilde \Gamma_h^2e^{\theta_h/T_h}
\eea
The cooling window is defined by the condition $0 < \theta_q < \theta_h\frac{T_q}{T_h}$.
When $\langle q\rangle \ll \Pi_{i=h,q,w}\tilde\Gamma_i$, the second term in the noise dominates, thus it is given by
\bea \langled q^2 \rangled \approx \left| \frac{e^{\theta_h/T_h}+e^{\theta_q/T_q}}{e^{\theta_h/T_h}-e^{\theta_q/T_q}} \right| \langle q\rangle = 
\left| \coth\left(\frac{\theta_h}{2T_h} - \frac{\theta_q}{2T_q}\right) \right| \langle q \rangle \equiv F(\theta_q) \langle q \rangle
\label{eq:q2app}
\eea
We now consider two refrigerators, A and B, with identical total gaps $\theta_h$, but different internal energies,  $\theta_{q_A}\neq\theta_{q_B}$. 
Correspondingly, each QAR supports the cooling  current $\langle q_k\rangle$  with the associated noise $\langled q_k^2\rangled$.
As for the lower bound, we would like to show that (recovering the energy gaps in the expressions for the currents and noises):
\bea
\eta^2 = \frac{(\theta_{q_A}\langle q_A \rangle + \theta_{q_B}\langle q_B \rangle)^2}{(\theta_{w_A}\langle q_A \rangle + \theta_{w_B}\langle q_B \rangle)^2} 
\leq \frac{\theta^2_{q_A}\langled q_A^2 \rangled + \theta^2_{q_B}\langled q_B^2 \rangled}{\theta^2_{w_A}\langled q_A^2 \rangled + \theta^2_{w_B}\langled q_B^2 \rangled} = \eta^{(2)} \label{eta2}
\eea
Using Eq. (\ref{eq:q2app}) and that $\theta_w = \theta_h - \theta_q$, we get
\bea
\frac{[\theta_{q_A}\langle q_A \rangle + \theta_{q_B}\langle q_B \rangle]^2}{[(\theta_h - \theta_{q_A})\langle q_A \rangle + (\theta_h-\theta_{q_B})\langle q_B \rangle]^2} 
\leq 
\frac{\theta^2_{q_A}F(\theta_{q_A})\langle q_A \rangle + \theta^2_{q_B}F(\theta_{q_B})\langle q_B \rangle}{(\theta_h-\theta_{q_A})^2F(\theta_{q_A})\langle q_A \rangle + (\theta_h-\theta_{q_B})^2F(\theta_{q_B})\langle q_B \rangle}.
\eea
After cross-multiplying, all the $\langle q\rangle^2$ terms vanish and we are left with 
\bea 
&&\sum_{i \neq k} \left(\theta^2_{q_i}(\theta_h-\theta_{q_k})^2F(\theta_{q_k}) + 2\theta_{q_i}\theta_{q_k}(\theta_h-\theta_{q_i})^2F(\theta_{q_i})\right) \langle q_i \rangle \\ 
\nonumber
&&\leq \sum_{i \neq k} \left(\theta^2_{q_k}(\theta_h-\theta_{q_i})^2F(\theta_{q_k}) + 2\theta_{q_i}^2(\theta_h-\theta_{q_i})(\theta_h-\theta_{q_k})F(\theta_{q_i}) \right)\langle q_i \rangle 
\eea
Since each current $\langle q \rangle$ can be set freely by setting the transition rates, the inequality is true if and only if each $\langle q \rangle$ part obeys the inequality:
\bea \theta^2_{q_A}(\theta_h-\theta_{q_B})^2F(\theta_{q_B}) + 2\theta_{q_A}\theta_{q_B}(\theta_h-\theta_{q_A})^2F(\theta_{q_A}) 
\leq \theta^2_{q_B}(\theta_h-\theta_{q_A})^2F(\theta_{q_B}) + 2\theta_{q_A}^2(\theta_h-\theta_{q_A})(\theta_h-\theta_{q_B})F(\theta_{q_A}). 
\nonumber\\
\label{eq:A7}
\eea
After additional manipulations we get
\bea \left[\theta_h^2\left(\theta_{q_A}^2-\theta_{q_B}^2\right) - 2\theta_h\theta_{q_A}\theta_{q_B}(\theta_{q_A}-\theta_{q_B})\right]F(\theta_{q_B})
 + \left[2\theta_{q_A}\theta_h(\theta_h-\theta_{q_A})(\theta_{q_B}-\theta_{q_A})\right]F(\theta_{q_A}) \leq 0 
 \label{eq:twoterms}
 \nonumber\\
\eea
The function $F(\theta)$ is positive, and it grows exponentially as $\theta$ approaches the asymptote at the edge of the cooling window,
$\theta_q = \theta_h\frac{T_q}{T_h}$, from below.
Therefore, if $\theta_{q_A} < \theta_{q_B} < \theta_{h}\frac{T_q}{T_h}$, 
the first term in Eq. (\ref{eq:twoterms}) dominates. It is  given by
\bea 
\theta_h(\theta_{q_A} + \theta_{q_B})(\theta_{q_A} - \theta_{q_B}) - 2\theta_{q_A}\theta_{q_B}(\theta_{q_A}-\theta_{q_B}) &\leq 0,
\eea
which is  reduced to
\bea
\theta_h(\theta_{q_A} + \theta_{q_B}) - 2\theta_{q_A}\theta_{q_B} &\geq 0. 
\eea
This is true since $\theta_h > \theta_{qA}$ and $\theta_h > \theta_{qB}$. 

Likewise, the second term in Eq. (\ref{eq:twoterms}) dominates in the opposite case,
if $\theta_{q_B} < \theta_{q_A} < \theta_{h}\frac{T_q}{T_h}$. We then check whether
\bea 
2\theta_{q_A}\theta_h(\theta_h-\theta_{q_A})(\theta_{q_B}-\theta_{q_A}) \leq 0,
 \eea 
 which is true, since the last term is negative and the rest are positive.

\section*{Appendix B: Lower bound on $\eta_2^{(2)}$ for three-level engines}
\setcounter{equation}{0}  
\renewcommand{\theequation}{B\arabic{equation}}
\label{app:2}


We consider here the performance of the three-level model as an {\it engine}, rather than a refrigerator.
We show that in the limit of small currents, at the edge of the engine's window, the lower bound for $\eta^{(2)}$ holds. This is to be contrasted with behavior observed for a thermoelectic engine, Appendix D, which shows violations to the lower bound in a corresponding domain.

The system acts as an engine when $\theta_h\frac{T_q}{T_h} < \theta_q < \theta_h$.
Heat is absorbed from the hot ($h$) bath and emitted at the $w$ bath as useful work, with leftout heat released at the $q$ bath. 
In this engine's regime, assuming the thermal noise dominates (since the current is small), we write
\bea
\langled q^2 \rangled =  -F(\theta_q)\langle q \rangle.
\eea
Recall that by our conventions, $\langle q\rangle$ is positive when flowing into the system, and thus it is negative when the system acts as an engine.
Furthermore, when the system operates as an engine, $F$ is positive and decreasing with $\theta_q$.

For an engine, the efficiency is the work current over the heat input from the hot bath. As usual, we set  the total gap size $\theta_h$  the same for both systems. Rewriting Eq. (\ref{eta2}) for the two engines gives
\bea
\eta^2 = \frac{(\theta_{w_A}\langle q_A \rangle + \theta_{w_B}\langle q_B \rangle)^2}{(\theta_{h_A}\langle q_A \rangle + \theta_{h_B}\langle q_B \rangle)^2} 
\leq \frac{\theta^2_{w_A}\langled q_A^2 \rangled + \theta^2_{w_B}\langled q_B^2 \rangled}{\theta^2_{h_A}\langled q_A^2 \rangled + \theta^2_{h_B}\langled q_B^2 \rangled} = \eta^{(2)}. \label{eta2engine}
\eea
Recall that $\langle q \rangle$ is a scaled measure, i.e., it counts the number of quanta exchanged, rather than the heat current itself.
Following  the same steps as in Appendix A, we end with Eq. (\ref{eq:A7}), but with $\theta_q\to \theta_w$, except in the $F$ functions, and $\theta_w \to \theta_h$.
As before, we chose without loss of generality  
only the $\langle q_A \rangle$ terms. The lower bound is true if 
\bea
2(\theta_{q_A} - \theta_{q_B})
(\theta_h-\theta_{q_A})F(\theta_{q_A}) \leq (\theta_{q_A} - \theta_{q_B})(2\theta_h -\theta_{q_A}-\theta_{q_B})F(\theta_{q_B}).
\eea
When $\theta_{q_A} > \theta_{q_B}$,
\bea
2(\theta_h-\theta_{q_A})F(\theta_{q_A}) \leq (2\theta_h -\theta_{q_A}-\theta_{q_B})F(\theta_{q_B})
\eea
which is true, using that $F$ is a decreasing function of $\theta_q$.
%
When $\theta_{q_A} < \theta_{q_B}$,
\bea
2(\theta_h-\theta_{q_A})F(\theta_{q_A}) \geq (2\theta_h -\theta_{q_A}-\theta_{q_B})F(\theta_{q_B}),
\eea
which also holds.

In Fig. \ref{fig:boundengine},  we search numerically for violations of the lower bound for an ensemble of three-level absorption engines, in a broad parameter space. As we are not able to identify such violations, we hypothesize that the three-level system acting as either a refrigerator or an engine satisfies the lower bound on $\eta_N^{(2)}$.

\begin{figure*}[htbp]
{\includegraphics[width=15.0cm] {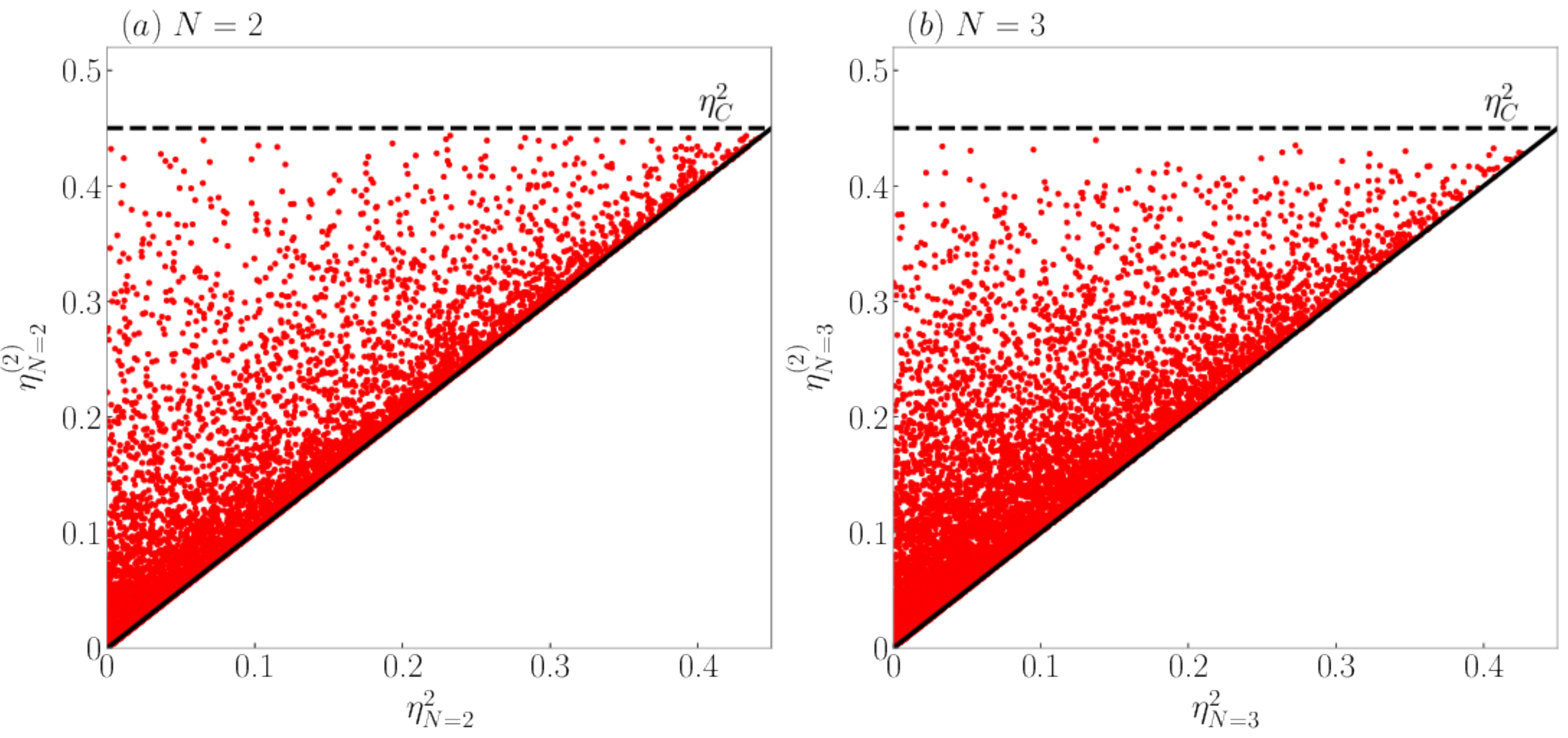} 
\caption{Demonstration that the lower bound $\eta_N^2\leq \eta^{(2)}_N$ is valid for the three-level absorption engine.
(a) Each dot corresponds to a pair of engines 
with random values for $\theta_q$ as well as coupling energies. 
For this pair, we calculate the total work and absorbed heat currents as well as their fluctuations. %
(b) Same calculations as in (a), but for a triplet of engines.
We used $\beta_q=0.4$, $\beta_h=0.2$,  and $ \beta_w=0.1$. The system-bath
coupling strengths are uniformly sampled from $0\leq\gamma_i\leq 1$, 
$\theta_q$ is sampled from the engine region.
  }}
\label{fig:boundengine}
\end{figure*}

\section*{Appendix C: Lower bound on $\eta_N^{(n)}$ for thermoelectric refrigerators in region I}

\setcounter{equation}{0}  
\renewcommand{\theequation}{C\arabic{equation}}
\label{app:3}

We now turn our attention to a pair of tight-coupling thermoelectric refrigerators, $A$ and $B$, with both $\epsilon_A$ and $\epsilon_B$ at the opposite end of the cooling window from those discussed in Section \ref{SecTE}. We refer to this as ``Region I", as depicted in Fig.~\ref{fig:single_ref}. In this region, $\langle q_k\rangle$, $k=A,B$, is small, vanishing when $\epsilon_k=\mu_L$, however, the charge current itself $\langle j_k\rangle$ remains finite and nonzero. We suppose that $\langle j_A\rangle \approx\langle j_B\rangle \approx\langle j_m\rangle|_{\epsilon_m = \mu_L}\equiv J$. Defining $u_k\equiv\epsilon_k - \mu_L$, we have that $\langle q_k\rangle = \langle j_k\rangle u_k \approx Ju_k$. The cooling efficiency of a single refrigerator is $\eta_k = u_k/\Delta\mu$, and the tight-coupling limit gives $\eta^{(2)}_k = (u_k/\Delta\mu)^2$.

Writing out the efficiency of the pair, we have
\be\label{eq:teII_eta}
    \eta_{N=2} \approx \frac{\langle q_A\rangle +\langle q_B\rangle }{2\Delta\mu J} \approx \frac{u_A + u_B}{2\Delta\mu} = \frac{\eta_A}{2}\bigg(1 + \frac{u_B}{u_A}\bigg).
\ee

Since Region I is {\it not} characterized by a small charge current, we cannot assume that the fluctuations in the charge or heat currents are given predominantly by thermal noise. 
However, we may echo our assumption above by supposing that $\langled j_A^2\rangled\approx\langled j_B^2\rangled\approx\langled j_m^2\rangled|_{\epsilon_m=\mu_L}\equiv S$. Then,  fluctuations in the cooling currents are given by $\langled q_k^2\rangled\approx Su_k^2$, and the ratio of cooling to work current fluctuations is given by
\be\label{eq:teII_eta2}
    \eta^{(2)}_{N=2} \approx \frac{u_A^2+u_B^2}{2\Delta\mu^2} =\frac{\eta_A^{(2)}}{2}\left[1 + \left(\frac{u_B}{u_A}\right)^2\right].
\ee
In Fig. \ref{fig:AppC} we demonstrate that  Eqs.~(\ref{eq:teII_eta}) and (\ref{eq:teII_eta2})
indeed provide a good approximation for the efficiency and the ratio of fluctuations for refrigerators in Region I. 

Squaring Eq.~(\ref{eq:teII_eta}), utilizing that $\eta_A^2 = \eta_A^{(2)}$, and comparing it to Eq.~(\ref{eq:teII_eta2}), we see that the lower bound, $\eta_{N=2}^2\leq\eta_{N=2}^{(2)}$, is equivalent to the inequality
\be
    \frac{1}{4}\bigg(1 + \frac{u_B}{u_A}\bigg)^2 \leq \frac{1}{2}\left[1 + \bigg(\frac{u_B}{u_A}\bigg)^2\right],
\ee
or, rearranging,
\be
    \bigg(\frac{u_B}{u_A} - 1\bigg)^2 \geq 0.
\ee
This is clearly satisfied for any choice of $u_A$ and $u_B$, so we conclude that the lower bound $\eta_{N=2}^2\leq\eta_{N=2}^{(2)}$ is satisfied for a pair of refrigerators in this regime.

We now extend this argument more generally to $n>2$ and $N>2$, by getting expressions for $\eta_N$ and $\eta_N^{(n)}$ of the ensemble analogous to Eqs.~(\ref{eq:teII_eta}) and (\ref{eq:teII_eta2}),
\bea
    \eta_N &=& \frac{\sum_{k=1}^N u_k}{N\Delta\mu},
    \nonumber\\
    \eta_N^{(n)} &=& \frac{\sum_{k=1}^Nu_k^n}{N\Delta\mu^n}.
\eea
Taking the $n^{th}$ root of $\eta_N^{(n)}$, we see that the lower bound, $\eta_N^n\leq\eta_N^{(n)}$, is equivalent to
\be 
    \frac{1}{N}\sum_{k=1}^N u_k \leq \bigg(\frac{1}{N}\sum_{k=1}^N u_k^n\bigg)^\frac{1}{n},
\ee 
which is always true as a result of the power mean inequality\cite{ineqHandbook}.

\begin{figure*}[t]
\includegraphics[width=13.0cm] {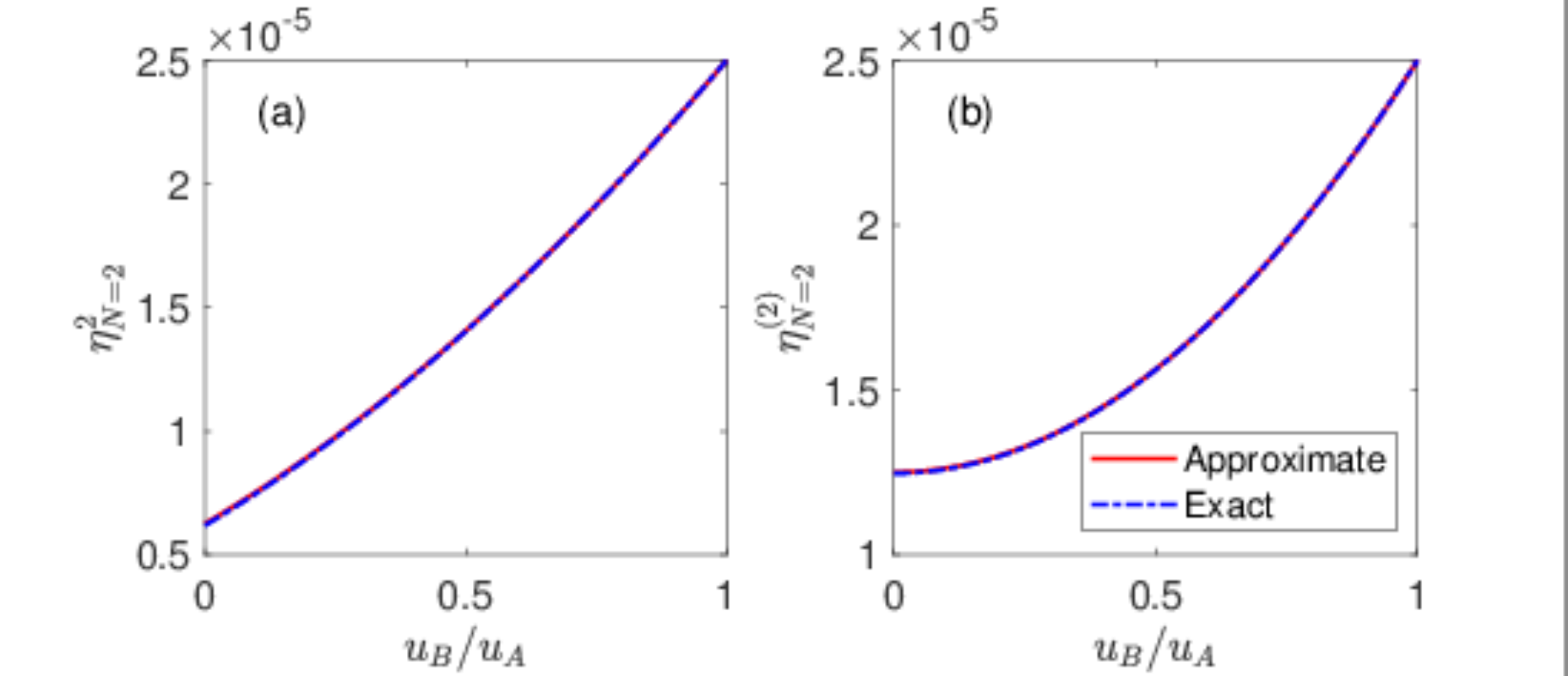}
\caption{Comparison of the approximate expressions for (a) $\eta_{N=2}^2$ and (b) $\eta_{N=2}^{(2)}$, given by Eqs.~(\ref{eq:teII_eta}) and (\ref{eq:teII_eta2}), respectively, with their exact values. We focus on the case $u_B<u_A$ to avoid redundancy. The good correspondence between approximate and exact values suggests that the approximations made are valid. Indeed, the lower bound is seen to be satisfied. $\beta_L = 2$, $\beta_R = 1$, $\mu_L = 1$, $\mu_R = -1$, $u_A = 0.01$, and $\Gamma_L = \Gamma_R = 1$.}
\label{fig:AppC}
\end{figure*}
%

\section*{Appendix D: Violation of the lower bound on $\eta_{N=2}^{(2)}$ for a pair of thermoelectric engines}

\setcounter{equation}{0}  
\renewcommand{\theequation}{D\arabic{equation}}
\label{app:4}

We consider here a pair of thermoelectric engines in the tight-coupling limit, far from equilibrium, operating in the regime of high $\epsilon_k$, for both $k=A,B$, and thus small charge current. Note that, due to the assumption of tight-coupling, this operational regime has no upper boundary with respect to $\epsilon_k$. Furthermore, we suppose that the magnitude of the charge current is considerably greater for engine $A$ than for $B$. Since, in this regime, charge current decreases with $\epsilon_k$, $\epsilon_B>\epsilon_A$, and thus, $\eta_A>\eta_B$. We will show that, under these assumptions, the lower bound on the ratio of fluctuations, $\eta_{N=2}^2\leq\eta_{N=2}^{(2)}$, must be violated.

In contrast to the refrigerator case, we now define the heat current with respect to the right (hot) lead--$\langle q_k\rangle = (\epsilon_k - \mu_R)\langle j_k\rangle$. We first write out the efficiency of the pair of engines,
\be
    \eta_{N=2} = \frac{\langle w\rangle}{\langle q\rangle} = \frac{\Delta\mu(\langle j_A\rangle + \langle j_B\rangle)}{(\epsilon_A - \mu_R)\langle j_A\rangle + (\epsilon_B - \mu_R)\langle j_B\rangle} = \eta_A\frac{1 + \frac{\langle j_B\rangle}{\langle j_A\rangle}}{1 + \frac{\epsilon_B - \mu_R}{\epsilon_A - \mu_R}\frac{\langle j_B\rangle}{\langle j_A\rangle}}.
\ee
Noting that $\langle j_B\rangle/\langle j_A\rangle\ll1$, we expand to first order in this ratio. After squaring, we have
\be\label{tc_engine_etasq}
    \eta_{N=2}^2 \approx \eta_A^2\left[1 + 2\bigg(1 - \frac{\epsilon_B - \mu_R}{\epsilon_A - \mu_R}\bigg)\frac{\langle j_B\rangle}{\langle j_A\rangle}\right] = 
    \eta_A^{(2)}\left[1 + 2\bigg(1 - \frac{\epsilon_B - \mu_R}{\epsilon_A - \mu_R}\bigg)\frac{\langle j_B\rangle}{\langle j_A\rangle}\right].
\ee
Next, we consider the fluctuations in the work and heat currents. The assumption of small currents allows us to neglect cotunneling processes, that is
contributions proportional to $(f_L(\epsilon_k) - f_R(\epsilon_k))^2$. Furthermore, we use the identity $f_L(\epsilon_k)\left[1 - f_R(\epsilon_k)\right] + f_R(\epsilon_k)\left[1 - f_L(\epsilon_k)\right] = [f_L(\epsilon_k) - f_R(\epsilon_k)]\coth[(-\Delta\beta\epsilon_k + \bar{\beta}\Delta\mu)/2]\equiv c_k[f_L(\epsilon_k) - f_R(\epsilon_k)]$ 
(note that $\Delta\beta=\beta_L - \beta_R<0$, $\bar \beta=(\beta_L+\beta_R)/2$) to express the charge current fluctuations for each engine in terms of the charge current itself,
\be
    \langled j_k^2\rangled = c_k\langle j_k\rangle.
\ee
%
\begin{figure*}[t]
\includegraphics[width=8.0cm] {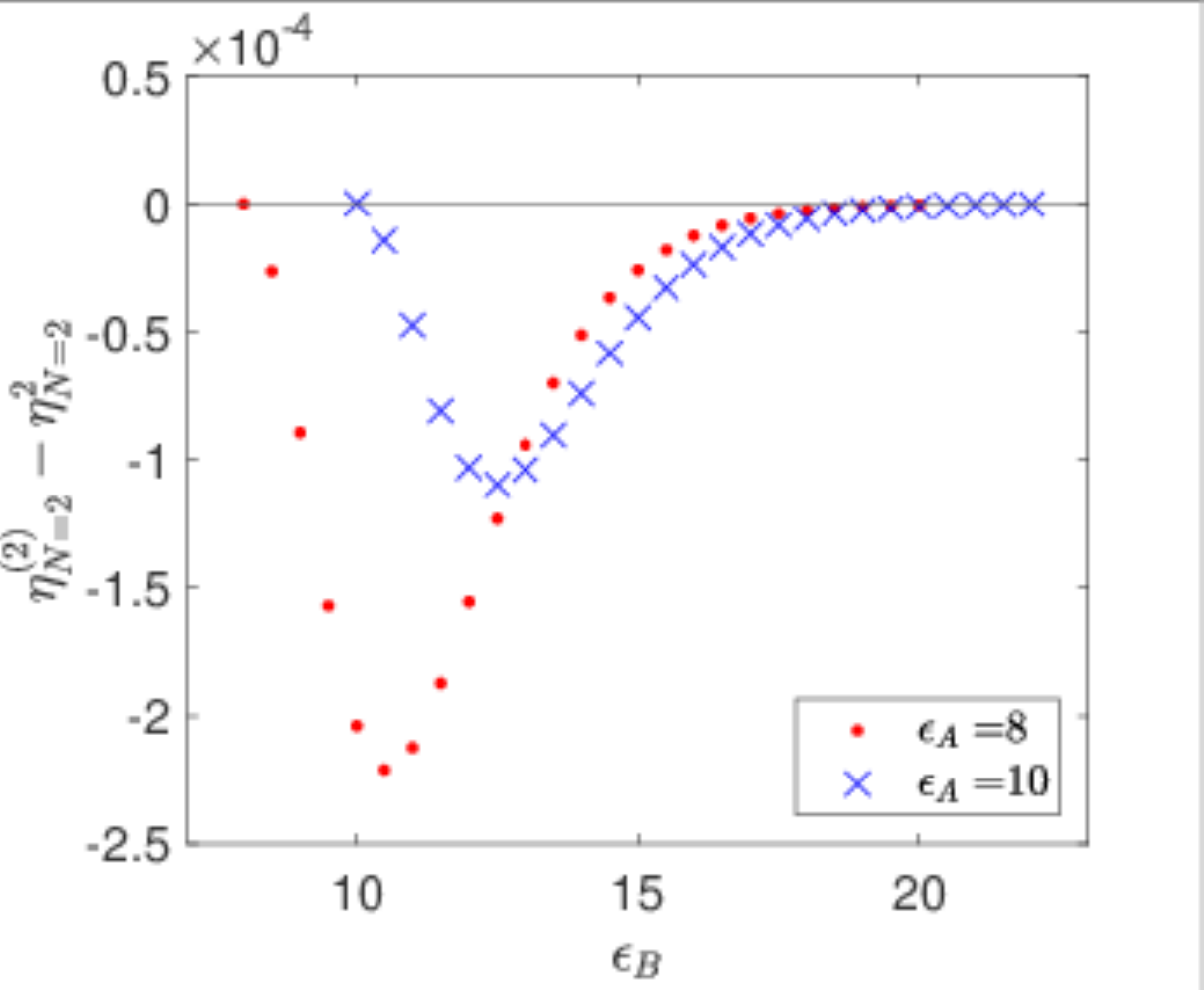} 
\caption{Violations of the lower bound observed for pairs of tight-coupling thermoelectric engines with sufficiently high $\epsilon_A$ and $\epsilon_B$, shown as a function of $\epsilon_B>\epsilon_A$ for two choices of $\epsilon_A$. Parameters are
$\beta_L = 2$, $\beta_R = 1$, $\mu_L = 1$, $\mu_R = -1$, and $\Gamma_L = \Gamma_R = 1$.}
\label{fig:AppD}
\end{figure*}
%
Then, the ratio of fluctuations for the pair is given by
\be
    \eta_{N=2}^{(2)} = \frac{\langled w^2\rangled}{\langled q^2\rangled} = \frac{\Delta\mu^2\left(c_A\langle j_A\rangle + c_B\langle j_B\rangle\right)}{(\epsilon_A - \mu_R)^2c_A\langle j_A\rangle + (\epsilon_B - \mu_R)^2c_B\langle j_B\rangle} = \eta_A^{(2)}\frac{1 + \frac{c_B\langle j_B\rangle}{c_A\langle j_A\rangle}}{1 + \frac{(\epsilon_B - \mu_R)^2c_B\langle j_B\rangle}{(\epsilon_A - \mu_R)^2c_A\langle j_A\rangle}}.
    \label{eq:D4}
\ee
In the engine regime, $\epsilon_B>\epsilon_A$ implies that $|c_B|<|c_A|$, so we do not have to worry that $c_B/c_A$ blows up and we may expand  Eq. (\ref{eq:D4}) to first order in $\langle j_B\rangle/\langle j_A\rangle$, giving
\be\label{tc_engine_eta2}
    \eta_{N=2}^{(2)} \approx \eta_A^{(2)}\left[1 + \bigg(1 - \bigg(\frac{\epsilon_B - \mu_R}{\epsilon_A - \mu_R}\bigg)^2\bigg)\frac{c_B\langle j_B\rangle}{c_A\langle j_A\rangle}\right].
\ee
Now, comparing Eqs.~(\ref{tc_engine_etasq}) and (\ref{tc_engine_eta2}), we find that the lower bound $\eta_{N=2}^2\leq\eta_{N=2}^{(2)}$ is equivalent to the inequality
\be 
    2\bigg(1 - \frac{\epsilon_B - \mu_R}{\epsilon_A - \mu_R}\bigg)\leq\frac{c_B}{c_A}\left[1 - \bigg(\frac{\epsilon_B - \mu_R}{\epsilon_A - \mu_R}\bigg)^2\right].
\ee
Recall that 
$c_k=\coth[(-\Delta\beta\epsilon_k + \bar{\beta}\Delta\mu)/2]$;
working far from equilibrium, we suppose that $\Delta\beta$ is sufficiently large that $\epsilon_A$ and $\epsilon_B$ meeting our above assumptions may be chosen such that both $c_A\approx -1$ and $c_B\approx-1$. Thus, $c_B/c_A\approx1$. This simplifies the necessary and sufficient condition for the lower bound to
\be
    2\bigg(1 - \frac{\epsilon_B - \mu_R}{\epsilon_A - \mu_R}\bigg)\leq\left[1 - \bigg(\frac{\epsilon_B - \mu_R}{\epsilon_A - \mu_R}\bigg)^2\right],
\ee
or, equivalently,
\be\label{tc_engine_contradiction}
    \bigg(\frac{\epsilon_B - \mu_R}{\epsilon_A - \mu_R} - 1\bigg)^2\leq 0.
\ee
Our assumption that $\epsilon_B>\epsilon_A$ renders Eq.~(\ref{tc_engine_contradiction}) a clear contradiction. Therefore, the set of assumptions considered here describes a pair of thermoelectric engines that \textit{must} violate the lower bound on the ratio of work to heat current fluctuations. We emphasize that this derivation relies on the thermoelectric engine operating far from equilibrium. In linear response, a small value for $\Delta\beta$ would conflict with the ability to achieve $c_B/c_A\approx1$ for any choice of $\epsilon_A$ and $\epsilon_B$ satisfying the rest of the assumptions. Thus, there is no contradiction with Ref.~\cite{Gerry}.

In Fig. \ref{fig:AppD} we provide supporting numerical evidence to the violation of the lower bound for thermoelectric engines operating in Region II. 

\end{widetext}

\end{document}

Add FCS as an Appendix? high cumulants?
Define Gamma, etc.

-Lower bound: proof for N>2
-high cumulants
-simulations
-Bijay's papers.
- Other models for QAR? ensemble?...
proving eta^3 = ()^3 for a single QAR

discuss expeeriments

add refs on QAR, fluctuations.
thermoelectic?

Need to reconcile Matthew's lower bounds expressions with the behavior of the QAR.
1. Test numerically Eqs. 9  and 10 for the QAR
2. Correct Eq. 18, the ensuing inequality, then Appendix A

perturbation treatments?!

Secs 1-3 Change to I and J 

QAR -->engine (violations?)
thermoelectric --> cooler (all satisfied?)

Refs: time dependent
eff fluctuations